\documentclass{aastex62}
\usepackage{float}
\usepackage{ulem}
\usepackage{breqn}
\usepackage{bm}
\usepackage{xcolor}

\newcommand\msunyr{\rm {\it M}_{\odot}\,yr^{-1}}
\newcommand\mum{$\mu$m}
\newcommand\mums{$\mu$m }
\newcommand\dotM{$\dot{M}$}

\begin{document}

\title{Determining Dust Properties in Protoplanetary Disks: SED-derived Masses and Settling With ALMA}
\author[0000-0002-3091-8061]{Anneliese M. Rilinger}
\affiliation{Department of Astronomy and Institute for Astrophysical Research, Boston University, 725 Commonwealth Avenue, Boston, MA 02215}

\author[0000-0001-9227-5949]{Catherine C. Espaillat}
\affiliation{Department of Astronomy and Institute for Astrophysical Research, Boston University, 725 Commonwealth Avenue, Boston, MA 02215}

\author[0000-0001-6060-2730]{Zihua Xin}
\affiliation{Department of Astronomy and Institute for Astrophysical Research, Boston University, 725 Commonwealth Avenue, Boston, MA 02215}

\author[0000-0003-3133-3580]{Álvaro Ribas}
\affiliation{Institute of Astronomy, University of Cambridge, Madingley Road, Cambridge CB3 0HA, UK}
\affiliation{European Southern Observatory (ESO), Alonso de Córdova 3107, Vitacura, Casilla 19001, Santiago de Chile, Chile}

\author[0000-0003-1283-6262]{Enrique Macías}
\affiliation{ESO Garching, Karl-Schwarzschild-Str. 2, 85748, Garching bei Munchen, Germany}

\author[0000-0002-1069-9834]{Sarah Luettgen}
\affiliation{Department of Astronomy and Institute for Astrophysical Research, Boston University, 725 Commonwealth Avenue, Boston, MA 02215}
\affiliation{Ann and H.J. Smead Department of Aerospace Engineering Sciences, University of Colorado at Boulder, 3775 Discovery Dr, Boulder, CO 80303}

\date{\today}
\correspondingauthor{Anneliese M. Rilinger}
\email{amr5@bu.edu}

\begin{abstract}
We present spectral energy distribution (SED) modeling of 338 disks around T Tauri stars from eleven star-forming regions, ranging from $\sim$0.5 to 10 Myr old. The disk masses we infer from our SED models are typically greater than those reported from (sub)mm surveys by a factor of 1.5-5, with the discrepancy being generally higher for the more massive disks. Masses derived from (sub)mm fluxes rely on the assumption that the disks are optically thin at all millimeter wavelengths, which may cause the disk masses to be underestimated since the observed flux is not sensitive to the whole mass in the disk; SED models do not make this assumption and thus yield higher masses. Disks with more absorbing material should be optically thicker at a given wavelength; which could lead to a larger discrepancy for disks around massive stars when the disk temperature is scaled by the stellar luminosity. We also compare the disk masses and degree of dust settling across the different star-forming regions and find that disks in younger regions have more massive disks than disks in older regions, but a similar degree of dust settling. Together, these results offer potential partial solutions to the ``missing'' mass problem: disks around T Tauri stars may indeed have enough material to form planetary systems, though previous studies have underestimated the mass by assuming the disks to be optically thin; these planetary systems may also form earlier than previously theorized since significant dust evolution (i.e., settling) is already apparent in young disks.
\end{abstract}

\keywords{protoplanetary disks - star formation}

\section{Introduction}
The population of confirmed exoplanets detected by Kepler, TESS, radial velocity surveys, and other methods, is diverse and growing \citep{dressing13, winn15}.  A complete picture of the formation process that can explain the origin of known exoplanets is not yet completely clear \citep[see for example the recent review by][]{drazkowska22}.  As the location of and source of material for planet formation \citep{williams11}, protoplanetary disks are a key piece of the puzzle.  Studying the properties of these disks can help inform and constrain various theories of planet formation, including the timescales over which the formation can occur. 

The Atacama Large Millimeter/submillimeter Array (ALMA) has been invaluable for studying protoplanetary disks.  In particular, large surveys of various star-forming regions performed with ALMA have provided information about disk masses \citep[see Table \ref{table:regs},][]{eisner18, vanterwisga20, vanterwisga22, anderson22} and dust substructures \citep[e.g.,][]{andrews18} for disks around T Tauri stars (TTS). The regions studied by these surveys span from approximately $\sim$0.5--10 Myr, allowing for comparisons of how disk properties evolve over time. 

\begin{deluxetable*}{c c c c c c c}
\tablecaption{(Sub)millimeter Surveys of Star-forming Regions Studied in this Work\label{table:regs}}
\tablehead{
\colhead{Region} & \colhead{Age} & \colhead{Approx. Dist.} & \colhead{Ref.} & N$_{total}$\tablenotemark{a} & N$_{det}$\tablenotemark{a} & N$_{sample}$ \\
\colhead{} & \colhead{Myr} & \colhead{pc} & \colhead{} & \colhead{} & \colhead{} & \colhead{}}
\startdata
$\rho$ Ophiuchus & 0.5 - 2 & 140 & \citet{cieza19, williams19} & 165 & 112 & 77\\
Taurus & 1 - 2 & 140 & \citet{andrews13, long19} & 98 & 65 & 25\\
L1641 & 1.5 & 428 & \citet{grant21} & 101 & 89 & 56\\
Cha II & 1 - 2 & 198 & \citet{villenave21} & 27 & 20 & 14\\
Lupus & 1 - 3 & 160 & \citet{ansdell16, ansdell18} & 87 & 62 & 41\\
Cha I & 2 - 3 & 180 & \citet{pascucci16} & 55 & 41 & 23\\
IC348 & 2 - 3 & 310 & \citet{ruiz-rodriguez18} & 131 & 36 & 24\\
Corona Australis & 3 & 160 & \citet{cazzoletti19} & 33 & 16 & 10\\
$\sigma$ Ori & 3 - 5 & 385 & \citet{ansdell17} & 88 & 37 & 31\\
$\lambda$ Ori & 5 & 400 & \citet{ansdell20} & 33 & 13 & 12\\
Upper Sco & 5 - 10 & 145 & \citet{barenfeld16} & 79 & 42 & 25\\
\enddata
\tablenotetext{a}{Objects in binary or multiple systems were excluded from these counts.}
\end{deluxetable*}

Two disk properties are especially informative for understanding the planet formation process: disk dust mass and dust settling. As the mass reservoirs for planet formation, the amount of dust and gas in a disk directly constrains the number and size of planets that could potentially form in that system. The review by \citet{bergin18} summarizes the various methods used to calculate disk masses \citep[see also][]{miotello22}. One of the most commonly applied methods calculates disk masses from millimeter-wavelength flux measurements. The disk masses deduced from the ALMA surveys listed in Table \ref{table:regs} are typically larger in younger regions (1--3 Myr) than in older regions \citep[see e.g.,][]{barenfeld16, vanderplas16, vanterwisga22}; these results can be used to constrain the evolution of protoplanetary disks and the timescale of planet formation.  Moreover, the estimated masses of the young disks are often too small to account for observed planetary systems \citep{manara18} or require nearly 100\% efficiency in the planet formation process \citep{mulders21}. A few possible solutions to this ``missing mass'' discrepancy have been offered. First, the masses obtained from millimeter observations may underestimate the actual disk masses \citep{ballering19, ribas20}.  Second, planet formation may occur rapidly, within the first few Myr of a disk's lifetime \citep[e.g.,][]{najita14}. Third, as possibly observed by \citet{ginski21}, the disk may accrete additional material (in significant amounts) from its surroundings, replenishing the disk \citep[e.g.,][]{throop08, kuffmeier17}; the amount of material we see in the disk would thus be only a fraction of the total material available for planet formation, alleviating the missing mass problem \citep{manara18}.

Dust settling is an important early step in the planet formation process. The dust grains in disks originate from the interstellar medium (ISM) as small ($\sim$0.25\mum{}) amorphous silicates. As the disk evolves, these dust grains collide with each other; if the collisions are inelastic the grains can grow to $\sim$micron sizes.  These larger grains feel a greater drag force as they decouple from the bulk gas motion and move through the gas in the disk. The grains therefore experience a net force towards the disk midplane and settle out of the disk atmosphere \citep{dullemond04}. Since this process takes time, it may be expected that older disks would be more settled than younger disks.  However, infrared (IR) observations have shown that this process is already occurring as early as $\sim$1 Myr \citep{furlan09, ribas17, grant18}. See the recent review by \citet{miotello22} for an overview of dust settling and disk vertical structure.

To build a complete picture of how disk masses and dust settling change over time -- and therefore how the planet formation process is unfolding -- the ideal disk sample would include as many disks as possible from star-forming regions spanning a wide range of ages.  Until recently, such a study has been difficult to implement, given the complexity and computational demands of the physical models used to determine disk properties. Fortunately, we now have access to machine-learning technology that can reproduce the results of a physical model in a fraction of the computational time: \citet{ribas20} trained an Artificial Neural Network (ANN) to replicate the output of the D'Alessio Irradiated Accretion Disk (DIAD) models \citep{diad98, diad99, diad01, diad05, diad06}. The ANN thus allows us to model hundreds of disks and perform a robust statistical analysis on their parameters.

In this work, we use this modeling framework to consistently model a large sample of protoplanetary disks in multiple star-forming regions in order to probe how disk mass and dust settling vary over time. In Section \ref{sample}, we present our sample of protoplanetary disks. Our modeling process and results are presented in Section \ref{results}, and we discuss our findings in Section \ref{discuss}. Finally, Section \ref{summary} provides a summary.

\section{Disk Sample}\label{sample}
We present here our sample of consistently-modeled T Tauri stars and their surrounding disks.  These objects are located in eleven star-forming regions, all of which have been studied at millimeter wavelengths: Ophiuchus \citep{cieza19, williams19}, Taurus \citep{andrews13, long19}, L1641 \citep{grant21}, Cha II \citep{villenave21}, Lupus \citep{ansdell16, ansdell18}, Cha I \citep{pascucci16}, IC348 \citep{ruiz-rodriguez18}, Corona Australis \citep{cazzoletti19}, $\sigma$ Ori \citep{ansdell17}, $\lambda$ Ori \citep{ansdell20}, and Upper Scorpius \citep{barenfeld16}. The results from the two recent ALMA surveys of Orion A \citep{vanterwisga22} and Serpens \citep{anderson22} are not included in our sample but will be discussed in a future work. See Table \ref{table:regs} for ages and distances for each region. Most of these surveys report to be complete down to the substellar limit (i.e., for spectral types $<$ M6).  However, the objects surveyed in L1641, Cha II, IC 348, and Upper Sco were selected based on 24\mums{} or 70\mums{} detections, meaning that these surveys are likely incomplete at low stellar masses (spectral types M4 and M5). Recent results using data from the Gaia surveys \citep{galli20, luhman20, galli21, grasser21, krolikowski21, esplin22} confirm the incompleteness of these samples. 

The total number of single Class II TTS observed in each region, N$_{total}$, is also reported in Table \ref{table:regs}. Objects in binary or multiple systems with separations $<$ 1000 au were not included because companion objects can alter the disk properties and influence disk evolution \citep{harris12, kounkel16, ruiz-rodriguez16, cox17, akeson19, barenfeld19, cazzoletti19, villenave21, rota22}. See the recent review by \citet{offner22} for a detailed discussion on the impact of binarity on disk systems.

In addition to being single objects, certain other criteria must be met for an object to be included in our sample.  We require the objects to have millimeter-wavelength photometry detections in order to constrain our SED models at long wavelengths. Though this requirement introduces a bias towards more massive disks since those disks are easier to detect in the millimeter, millimeter detections are crucial for constraining the model fit at long wavelengths and therefore the disk mass. The number of single Class II objects detected in the millimeter in each region, $N_{det}$ is reported in Table \ref{table:regs}. We also require objects to have significant photometric and/or spectroscopic coverage in the infrared to constrain the rest of the SED.  This requirement excludes all disks in NGC2024 \citep[0.5 Myr, 414 pc;][]{vanterwisga20} and Orion \citep[1 Myr, 400 pc;][]{eisner18}, despite their ALMA coverage, since they lack sufficient photometry points.  

No transitional disks (TDs) are included in our sample, since their inner cavities can affect the shape of an object's SED \citep[see for example the review by][]{andrews20}; this effect on the SED is not accounted for in our modeling framework, so we exclude these objects to avoid that source of uncertainty. TDs were determined by the shape of their SED \citep{cieza10, espaillat12, luhmanmamajek12, mauco16, vandermarel16, grant18} or through resolved imaging \citep{isella10, andrews11, canovas15}. As noted by \citet{andrews20}, the current catalog of very high-resolution observations in which substructure could be detected is biased towards larger, brighter disks around more massive hosts. Excluding TDs from our sample may preferentially exclude more massive disks and introduce a bias into our sample.

After applying these selection criteria, our modeling sample consists of 383 disks around T Tauri stars. Of these objects, 338 are used for our analysis; Section \ref{fitresults} describes our reasons for excluding 45 objects from the analysis. Table \ref{table:regs} lists the number of objects in our sample used for analysis per region, N$_{sample}$.

\section{Analysis and Results}\label{results}

\subsection{SED Models}\label{SEDmodels}
In order to determine disk properties for each of the TTS objects in our sample, we fit models to each object's SED.  SEDs were constructed using photometry points spanning visual to millimeter wavelengths, obtained using the Vizier catalog access tool \citep{vizier}.  We include photometry from \textit{Gaia} DR3, the American Association of Variable Star Observers Photometric All-Sky Survey \citep[AAVSO;][]{aavso2, aavso1}, the Guide Star Catalog \citep[GSC2.3;][]{gsc}, 2MASS, WISE, \textit{Spitzer} (labeled as c2d in the figures), \textit{Herschel}, and any available millimeter photometry (usually ALMA, but also including the Submillimetre Common-User Bolometer Array on the James Clerk Maxwell Telescope \citep[SCUBA;][]{scuba} and the Submillimeter Array \citep[SMA;][]{sma}). After visual inspection of the SED, we removed a few photometry data which were in obvious disagreement with the rest of the SED and which had apparent contamination in their images. We also include low-resolution (R $\sim$ 60–130) spectra from the InfraRed Spectrograph (IRS) on the \textit{Spitzer Space Telescope} \citep{houck04} for the 187 objects in our sample that were observed with this instrument. In order to counteract uneven weighting in the fitting process due to the much larger number of points in a spectrum than in the photometry for a given object, we bin the spectrum into a smaller number of points. Spectrum points are binned according to the nearest wavelength at which our model fluxes are calculated, yielding approximately 10-20 binned points, depending on the wavelength span of the spectrum.

\subsubsection{The Artificial Neural Network}\label{ann}
Previous studies \citep[e.g.,][]{rubinstein18, macias18, rilinger19, rilinger21} used the DIAD code \citep{diad98, diad99, diad01, diad05, diad06} to successfully model disk SEDs.  However, these models are computationally expensive, and the ability to perform robust statistical analysis is limited.  In this work, we use the ANN created by \citet{ribas20} to model the disk SEDs. The ANN was trained on tens of thousands of DIAD models, and can mimic DIAD outputs in a fraction of the computational time. A second ANN, ANN$_{diskmass}$ is employed to determine the disk mass for each object based on the best-fit model parameters. Neither the DIAD models nor the ANN include disk mass as a free parameter; rather, disk masses are inferred from the other best-fit parameters. See the description of $\alpha$ and \dotM{} below as well as \citet{ribas20} for details.

DIAD takes various disk and stellar parameters as inputs to calculate the emission from the system.  The dust in the disk is assumed to be distributed in two populations: smaller dust grains in the atmosphere of the disk and larger dust grains in the disk midplane.  These populations are parameterized by the maximum dust grain size in each population (i.e., $a_{\rm max, upper}$ and $a_{\rm max, midplane}$, respectively).  The amount of dust settling that has occurred is represented by the dimensionless parameter $\epsilon$ as described by \citet{diad06} and defined as follows. We assume the grain size distributions and the dust-to-gas mass ratio ($\zeta$) to be constant in radius and vary only with height above and below the disk midplane. In order to keep the total $\zeta$ constant while still allowing for settling of large grains, $\zeta_{small}$, the dust-to-gas ratio of small grains in the atmosphere (with sizes up to $a_{\rm max, upper}$) decreases while $\zeta_{big}$, the dust-to-gas ratio of larger grains in the midplane (with sizes up to $a_{\rm max, midplane}$) increases. The settling parameter $\epsilon$ is thus defined such that

\begin{equation}
\epsilon = \zeta_{small}/\zeta_{std}
\end{equation}

\noindent where $\zeta_{std}$ is the standard assumed dust-to-gas mass ratio of 0.01. Therefore, lower $\epsilon$ values correspond to larger $\zeta_{big}$, implying more settled disks.


Disk viscosity is characterized by $\alpha$, following \citet{shakura73}.  In the DIAD models, $\alpha$ and the mass accretion rate $\dot{M}$ are the two main parameters that set the disk mass.  The surface mass density $\Sigma$ is proportional to \dotM $\alpha^{-1}$; integrating $\Sigma$ over the disk radius yields disk mass. For a fixed mass accretion rate, smaller values of alpha thus correspond to larger disk masses, and vice versa. As discussed by \citet{ribas20}, $\alpha$ and \dotM{} can therefore be correlated, but they also have other smaller effects on the SED that can help constrain them. We allow alpha to vary between $10^{-4}$ and $10^{-1}$. These values are consistent with those derived by \citet{rafikov17} and \citet{ansdell18}; these works calculated $\alpha$ using measured mass accretion rates and disk sizes. ALMA molecular line observations generally support low turbulence values \citep[on the order of 10$^{-4}$ -- 10${-3}$][]{flaherty15, pinte16, flaherty17}, though some objects show higher values \citep[$\sim$ 0.01 -- 0.1][]{flaherty18, flaherty20}. We vary \dotM{} between $10^{-10}\ \msunyr{}$ and $10^{-6.5}\ \msunyr{}$; this is consistent with typical accretion rates reported for Class II disks \citep[e.g.,][]{valenti93, hartigan95, gullbring98, ingleby13, manara16a, simon16}. 

We note that DIAD calculates the radial and vertical disk density and temperature structure self-consistently. In general, the disk emission at a given wavelength will depend on the optical depth along the line of sight and the disk temperature structure. When the disk emission is optically thick, the temperature at the surface where the line-of-sight optical depth is unity will be the dominant component. As such, even when a disk is optically thick from the optical to the mm wavelengths, DIAD can still provide strong constraints on various disk parameters.

The geometry of the disk is also important for fitting the SED, so the disk radius $R_{\rm disk}$ and inclination $i$ are also included as input parameters.  Furthermore, the inner edge of the disk is an important source of disk emission, since the inner disk wall is assumed to be directly irradiated by the central star.  The location of the inner wall is defined by the dust sublimation temperature, $T_{wall}$, and the height of the wall is scaled by a factor $z_{\rm wall}$.

In addition to the nine disk parameters described above, the stellar temperature $T_*$, radius $R_*$, and mass $M_*$ are also important input parameters for DIAD (and therefore ANN).  We ensure that we only allow for combinations of stellar parameters that are consistent with models of stellar evolution, following \citet{ribas20}. We incorporate an $Age_*$ parameter into our fitting process (see Section \ref{mcmc}); combined with $M_*$, we use this parameter to calculate $T_*$ and $R_*$ based on the MESA Isochrones and Stellar Tracks \citep[MIST;][]{paxton11, paxton13, paxton15, dotter16, choi16}.  The output $T_*$ and $R_*$ are therefore consistent with the input $M_*$, and all three parameters form a consistent set that can be used as input for DIAD. We generally find strong agreement between the $T_*$ and $R_*$ calculated by our models and the values reported in the literature. See Appendix \ref{app:tstarrstar} for a comparison with literature values. We include these stellar parameters in our fitting procedure in order to allow for a complete Bayesian analysis of the rest of the parameters. As discussed in \citet{ribas20}, the stellar parameters derived in this fitting process are not intended to be precisely accurate values, as would be expected from a complete treatment of photospheric spectra. Rather, our values of $M_*$, $Age_*$, $T_*$, and $R_*$ are used to ensure that the models are internally consistent and to account for the uncertainties they produce in other parameters.

Finally, once an SED has been calculated using these input parameters, we redden the output SED to the appropriate extinction value $A_v$ using the \citet{mcclure09} extinction law.  We determine the $A_v$ values using the Markov Chain Monte Carlo (MCMC) fitting process described below in Section \ref{mcmc}. The reddened SEDs are ultimately scaled to the object's distance, which is also fit with the MCMC. \textit{Gaia} parallaxes \citep{gaia16, gaia18, gaia21} are used as priors where possible; the approximate distances to each region given in Table \ref{table:regs} are used for the 42 objects without \textit{Gaia} measurements.

\subsubsection{MCMC SED Modeling}\label{mcmc}
Including all of the free parameters described in the previous section, DIAD takes 13 input parameters.  In addition to these parameters, we include four additional parameters to account for photometric uncertainties and outliers. One of these is the free parameter $f$ by which we scale uncertainties in flux density to account for possible systematic uncertainties, as follows:

\begin{equation}
    s_n^2 = \sigma_{n,model}^2+\sigma_{n,obs}^2+f^2\,y_{n,obs}^2\quad
\end{equation}

\noindent where index n corresponds to each measurement, $\sigma_{n,model}$ is the adopted 10$\%$ uncertainty for the ANN prediction at the same wavelength, $\sigma_{n,obs}$ is the uncertainty in the flux for the measurement, $y_{n,obs}$ represents the observed fluxes of the measurement. The other three are parameters in a mixture model used to account for possible photometric outliers.  Following \citet{hogg10}, $P_{\rm out}$ is the probability that any photometry point is an outlier; $y_{\rm out}$ and $V_{\rm out}$ are the mean and variance of the outliers, respectively. Our model thus has a total of 17 free parameters.

In order to sufficiently probe the large parameter space and to estimate posterior distributions for the model parameters, we implement a Bayesian analysis framework in the form of a Markov Chain Monte Carlo (MCMC) fitting process.  We adopt the likelihood functions used by Xin et al. (in press), and the \texttt{ptemcee} \citep{vousden16} parallel-tempering version of the more commonly used \texttt{emcee} MCMC code \citep{foreman-mackey13}.  Following Xin et al. (in press), we separate the photometry points into two categories: ``critical'', which includes the 2MASS and millimeter-wavelength points, and ``general'', which includes all other photometry points. This categorization allows us to ascribe a higher weighting to the 2MASS and ALMA photometry; fitting these points is crucial for accurately scaling the stellar photosphere and calculating the disk mass, respectively. We use four standard Gaussian likelihood functions, one each for the critical photometry data, the spectral data, $T_*$, and $R_*$, given as:

\begin{equation}
\text{$\mathcal{L}$}_{n,data}= \frac{1}{\sqrt{2\pi (s_n^2)}} \exp{\biggl(-\frac{(y_{n,data} - y_{n,{\rm model}})^2}{2(s_n^2)}\biggr)}
\end{equation}

\noindent where $data$ is the data set (critical photometry, IRS spectrum, $T_*$, or $R_*$), index n corresponds to each measurement, $y_{n,x}$ represents the observed values, $y_{n,{\rm model}}$ represents the values predicted by ANN, and $s_n$ is the uncertainty scaling described above. We use the mixture model outlined above for the general photometry data (all other photometry points):

\begin{dmath}
\text{$\mathcal{L}$}_{n,gen}= \frac{1-P_{\text{out}}}{\sqrt{2\pi(s_n^2)}} \exp{\biggl(-\frac{(y_{n,gen} - y_{n,{\text{model}}})^2}{2(s_n^2)}\biggr)}
 + \frac{P_{\text{out}}}{\sqrt{2\pi [s_n^2 + V_{\text{out}}^2]}} \exp{\biggl(-\frac{(y_{n,gen} - y_{\text{out}})^2}{2 [s_n^2 +
    V_{\text{out}}^2]}\biggr)}
\end{dmath}

\noindent where $y_{n,gen}$ represents the observed fluxes of the photometry points.  The overall likelihood function is thus:

\begin{dmath}
\ln\mathcal{L} = \sum_{n=1}^{N_{\rm gen}}\ln\mathcal{L}_{n,gen} + \sum_{n=1}^{N_{\rm crit}}\ln\mathcal{L}_{n,crit} 
+ \sum_{n=1}^{N_{\rm spect}}\ln\mathcal{L}_{n,spect} + \ln\text{$\mathcal{L}$}_{T_{\rm eff}} + \ln\text{$\mathcal{L}$}_{R_*}
\end{dmath}

We adopt flat priors for most parameters, except those for which we use reported values in the literature.  The dust sublimation temperature $T_{wall}$ has a Gaussian prior centered around the typically-assumed value of 1400 K with a standard deviation of 50 K.  Gaussian priors are also used for parallax and $M_*$ for objects with \textit{Gaia} parallaxes, and reported $M_*$ values, respectively, using their values as the centers and their uncertainties as the standard deviations.  For objects with reported disk inclinations, we adopt a Gaussian prior, truncated at the limits of 0 and 70 degrees of inclination (the range of values on which the ANN was trained).  The maximum dust grain sizes $a_{\rm max,\ upper}$ and $a_{\rm max,\ midplane}$ are both fit with Jeffreys priors \citep{jeffreys46, jeffreys61}, within their limits of 0.25 - 10 \mums and 100 - 10$^4$ \mum, respectively.  All other parameters are assumed to have flat, uniform priors, within their allowed ranges of values (see Table C.1 in \citet{ribas20} for the ranges).

We run \texttt{ptemcee} with three temperatures and 102 walkers (the minimum number of required walkers for three temperatures and 17 parameters) for 2 $\times$ 10$^5$ steps. This corresponds to $\sim$2000 times the autocorrelation time. We visually inspected the chain to confirm that we had achieved convergence, and discard the first 10$^5$ steps based on this inspection. The remaining 10$^5$ steps are therefore converged and represent $\sim$1000 times the autocorrelation time. Since using all 10$^5$ steps would be computationally expensive in our analysis, we sample 10$^4$ steps from these converged steps, which is still sufficient to explore the parameter space.  These 10$^4$ steps were used to construct posterior distributions for the properties of interest, and to obtain the input parameters needed for ANN$_{diskmass}$ to calculate the corresponding distribution of disk masses. To determine the best-fit parameters, we take the median of the $10^4$ steps for each parameter, and report the $1\sigma$ value for the uncertainties.

\subsection{Model Results}\label{fitresults}
\begin{figure*}
\epsscale{1.2}
    \plotone{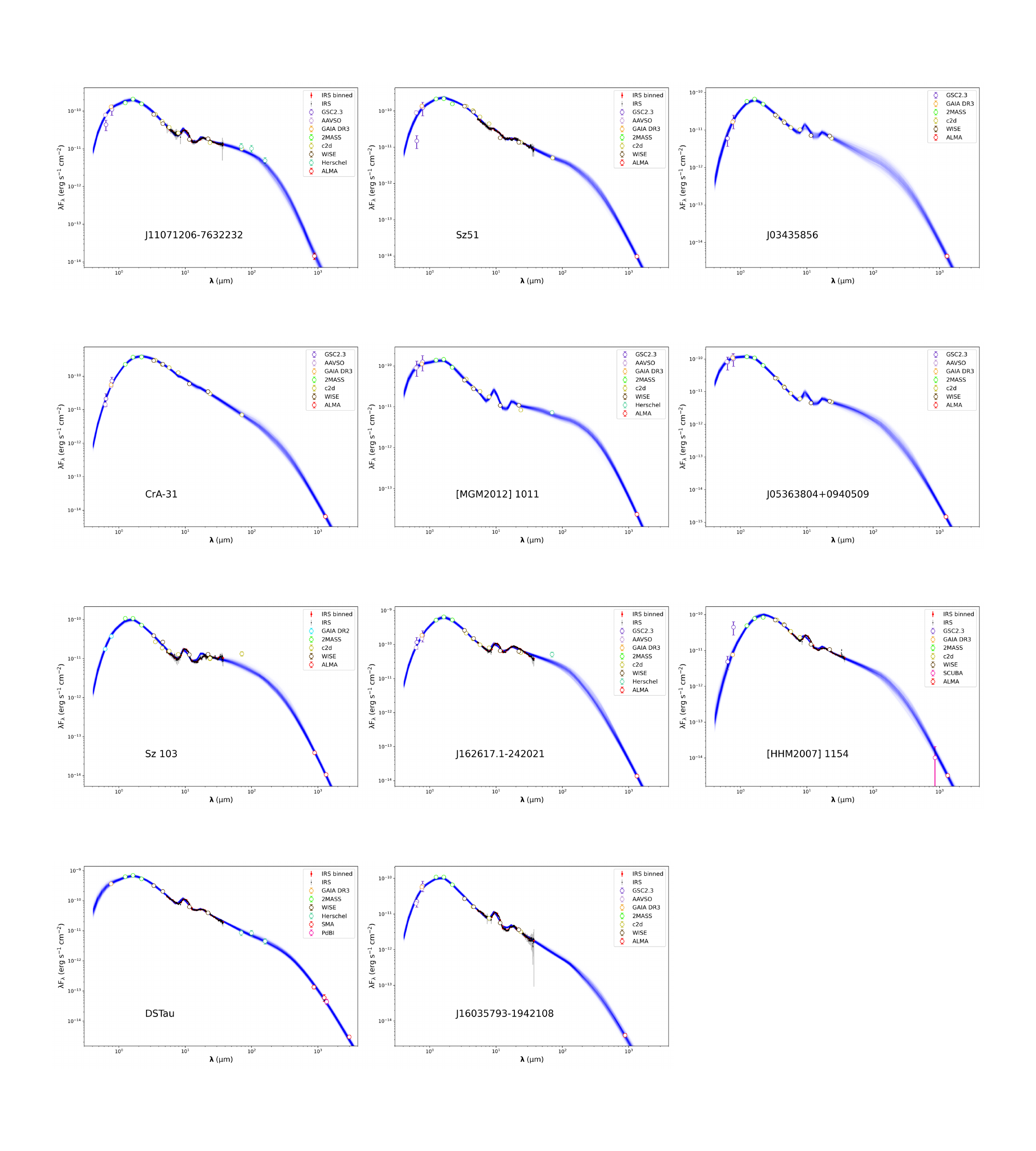}
    \caption{The observed SED and models for one example disk from each region. Results of the modeling process are shown by randomly selecting 1000 models (blue lines) from the posterior distributions. Instrument abbreviations are defined in Section \ref{SEDmodels}. SEDs and models for all objects are available in an online figure set. Some objects have upper limit photometry points; these are represented by downward-facing triangles in the online figure set. We note that the photometry has not been dereddened; we redden the models as described in Section \ref{ann}.}
    \label{fig:exampleSED}
\end{figure*}

Using the MCMC fitting process described above, we obtain SED models for each of the 383 objects in our sample.  In order to fully visualize the resulting fits for each object, we randomly sample 1000 steps from the converged MCMC chain and plot the corresponding models along with the photometry and, where available, an IRS spectrum. Some example SED fits are shown in Figure \ref{fig:exampleSED}; SED fits for the other objects are available in an online figure set. The corner plots for all objects are available on \dataset[Zenodo]{https://zenodo.org/record/7235076}.

\begin{figure*}
\epsscale{1.2}
    \plotone{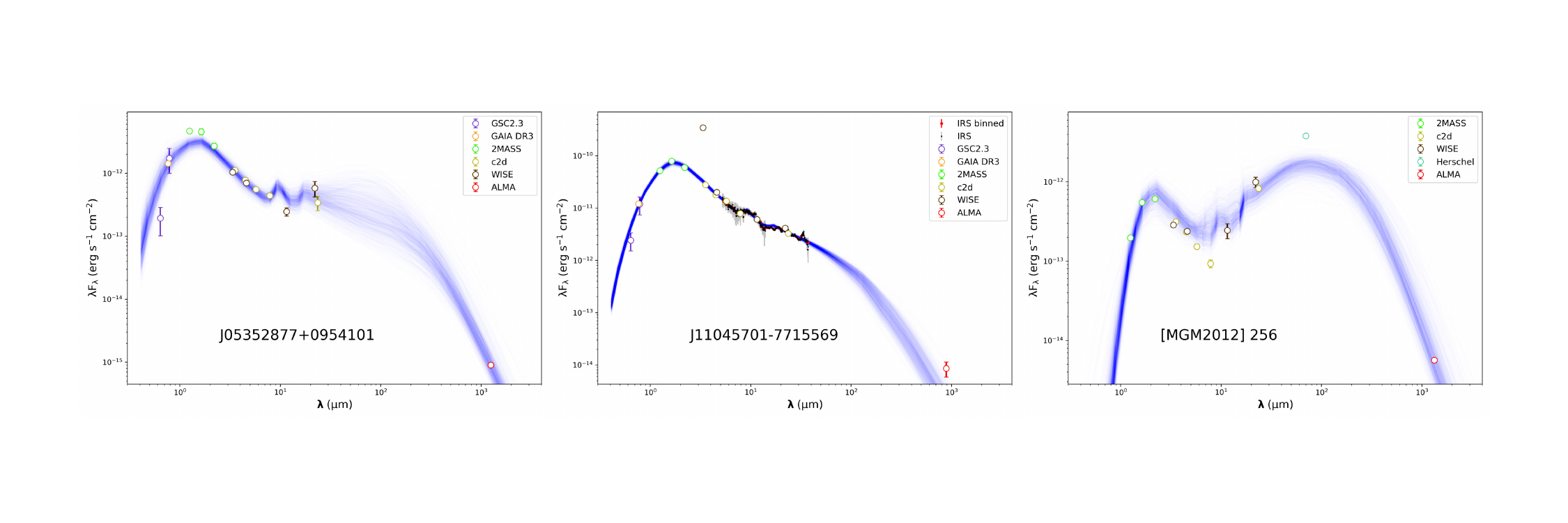}
    \caption{Three examples of unsuccessful SED fits. In the left panel, the models fail to fit the J, H, and K band photometry at the peak; in the middle panel, the models fail to fit the millimeter photometry; in the right panel, the shape of the SED indicates that this object might be highly inclined (i.e., viewed nearly edge-on), which would require a scattered-light model component to be fit well.}
    \label{fig:badSED}
\end{figure*}

After visual inspection, we determined that we obtained successful SED models, in which the models reproduce the observed SED, for 338 of the 383 objects in our sample, approximately 88\%. Three examples of unsuccessful fits are shown in Figure \ref{fig:badSED}. The unsuccessful fits can be attributed to a few explanations. First, TTSs are well-known to be variable on timescales ranging from minutes to decades \citep{siwak18}. This can explain discrepancies between photometric observations that were taken at different times. Since our model does not account for variability, differences between our model and the photometry -- particularly at shorter wavelengths, where variability is more pronounced -- are to be expected in some cases. Discrepancies between our model and near-IR photometry can also occur for disks around cool, very low mass stars, since the \citet{pecaut13} colors used by DIAD are more uncertain for late spectral types (see for example, the left panel in Figure \ref{fig:badSED}).  

Another potential source of disagreement is our assumption regarding dust composition. The DIAD models upon which the ANN is based assume a standard mixture of olivine silicates and graphite and adopt standard opacities calculated from Mie theory. The optical constants are taken from \citet{dorschner95} and \citet{draine84} for the silicates and graphite, respectively. DIAD, and therefore the ANN, do not fit spectral features in detail, but generally replicate the shape and size of the spectrum. In cases where our model does not perfectly replicate the IRS spectrum, the disk may have a different dust composition than assumed in our model.  

Millimeter photometry points with large uncertainties can also cause issues with the SED fit. As mentioned in Section \ref{sample}, millimeter photometry is crucial for fitting the SED at long wavelengths, especially since in many cases, the millimeter point is the only photometry point beyond 24 \mum{}.  Points with larger uncertainties are given lower weighting in the MCMC fitting process; if the millimeter uncertainties are too large, the model can miss that point (see for example the middle panel of Figure \ref{fig:badSED}). Future millimeter observations of these objects at higher sensitivities may result in lower uncertainties and therefore better-constrained model fits.

Finally, the geometry and configuration of the system may add extra challenges to fitting the SED. If a disk is highly inclined (i.e., viewed nearly edge-on), a scattered light model component would be necessary for an accurate fit (see Figure \ref{fig:badSED}, right panel); though DIAD can include scattered light, this feature is not yet available with the ANN. Additionally, though we attempted to exclude all objects with known companions and/or disk substructures, it is likely that some objects in our sample have companions or substructures that are yet undetected. Such a system may affect the SED in ways that are not accounted for by the DIAD models.  Thus, some amount of unsuccessful SED fits is expected.

Despite these challenges, we obtain successful model fits for 338 of the 383 objects in our sample.  Moving forward, we consider only the 338 objects with satisfactory model fits.

\section{Discussion}\label{discuss}
\subsection{The Role of Optical Depth on Derived Disk Masses}\label{masscomp}
In this section, we compare the disk masses inferred from our modeling process to the masses previously reported in the literature.  Typically, the following equation is used to convert an observed millimeter-wavelength flux to a dust mass \citep{hildebrand83}:

\begin{equation}\label{eq:mass}
    M_d = \frac{F_\nu d^2}{\kappa_\nu B_\nu(T_d)}
\end{equation}

\noindent where $M_d$ is the mass of dust in the disk, $F_{\nu}$ is the millimeter flux density, $\kappa_{\nu}$ is the dust opacity coefficient at frequency $\nu$, and $B_{\nu}$($T_d$) is the Planck function evaluated at dust temperature T$_d$.  This equation relies on some assumptions, namely, that the disk is isothermal and optically thin at the reference frequency.  Furthermore, assumptions must be made regarding the value of $T_d$ (whether it is constant for all disks, or if it scales with luminosity), and the value of $\kappa_{\nu}$ and how it scales with frequency.  Finally, many of the previous millimeter surveys of disks were published prior to the \textit{Gaia} data releases, and therefore had to rely on constant distance assumptions for all objects, as opposed to precise measurements.

\begin{figure*}
    \gridline{\fig{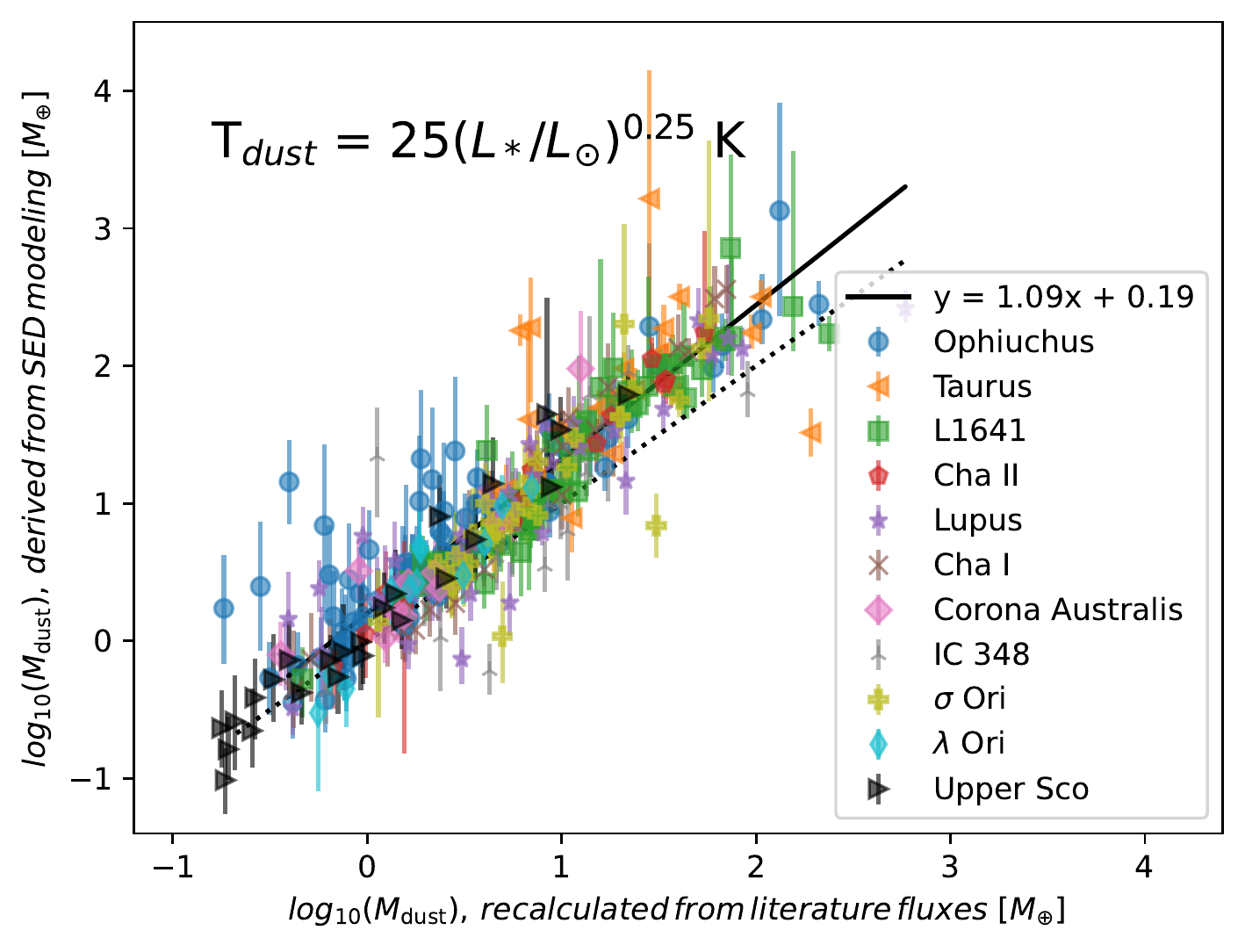}{0.5\textwidth}{(a)}
            \fig{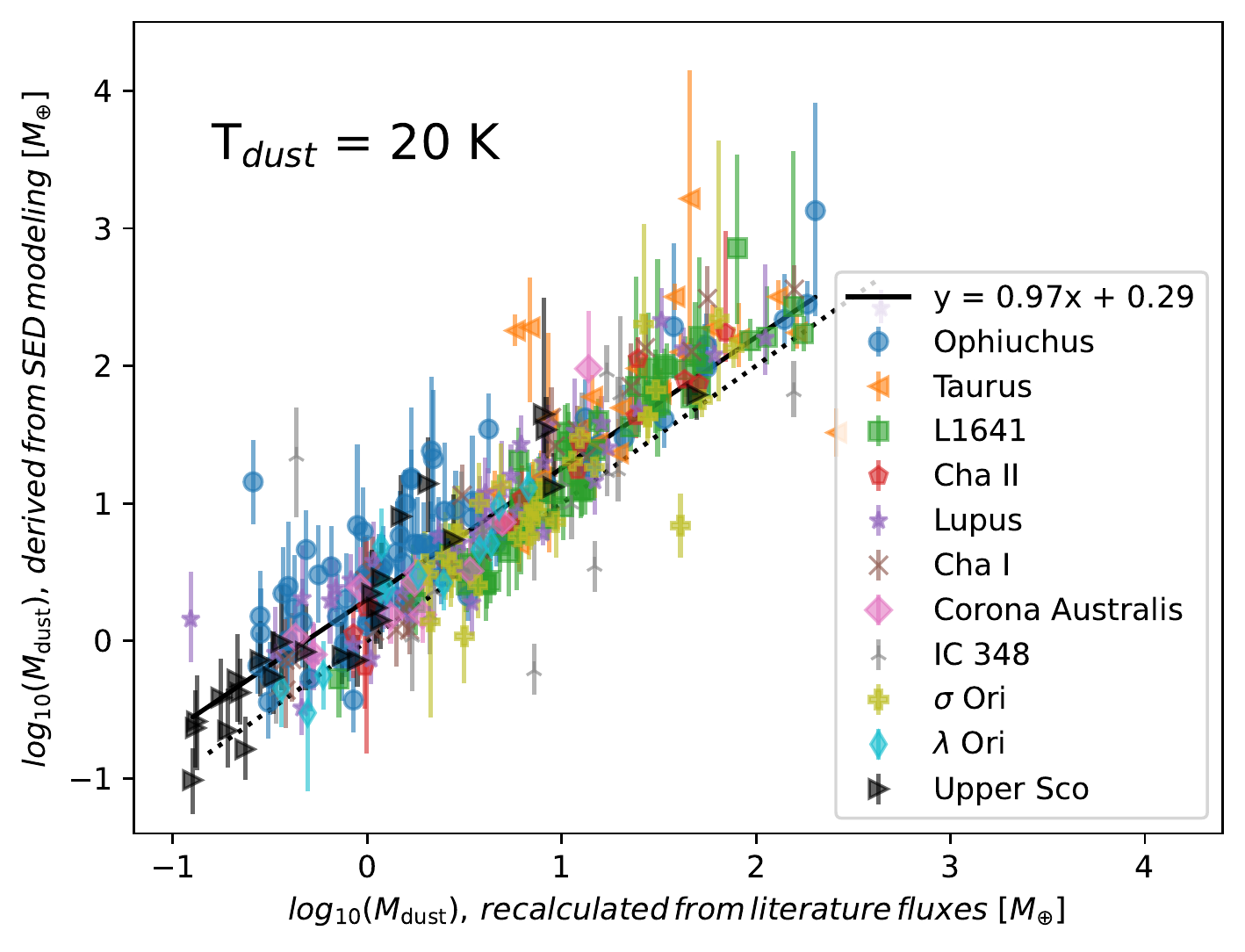}{0.5\textwidth}{(b)}}
    \caption{Comparison between our disk masses from SED modeling and disk masses calculated from single millimeter-wavelength fluxes. In Panel (a), we scale the dust temperature according to the stellar luminosity (calculated from our modeled T$_*$ and R$_*$ values), as in \citet{andrews13}.  In Panel (b), we assume a constant dust temperature of 20 K. In both panels, the black solid line represents the best-fit linear relationship and the black dotted line represents a one-to-one relationship. The linear fits presented here can be used to correct disk masses calculated using Equation \ref{eq:mass}. Uncertainties in the fitting parameters are provided in the text.}
    \label{fig:massrecalc}
\end{figure*}

Different studies chose different assumptions for the terms in Equation \ref{eq:mass}, so to ensure that we made a fair comparison, we recalculated all of the disk masses using Equation \ref{eq:mass} according to a consistent set of assumptions.  All distances were calculated from \textit{Gaia} parallaxes, where possible (the distances in Table \ref{table:regs} were used for objects without \textit{Gaia} data). In previous studies, dust temperature was assumed either to be a constant value \citep[typically 20 K;][]{ansdell16, ansdell17, ruiz-rodriguez18, cazzoletti19, williams19, ansdell20} or to vary with stellar luminosity:

\begin{equation}\label{eq:td}
    T_d = 25 (L_* / L_{\odot})^{0.25} K
\end{equation}

\noindent as in \citet{andrews13}, \citet{barenfeld16}, \citet{pascucci16}, and \citet{villenave21}. We set $\kappa_{\nu}$ to 2.3 cm$^2$ g$^{-1}$ at 230 GHz \citep{beckwith90}. A common assumption is that $\kappa_{\nu}$ scales with frequency as $\nu^{1.0}$ \citep[i.e., the power law index $\beta$ = 1.0,][]{ansdell16, ansdell17, cieza19, ansdell20, grant21}. Other works \citep[i.e.,][]{pascucci16, barenfeld16} use $\beta$ = 0.4. We set $\nu$ to the frequency of the ALMA continuum observations: 225 GHz (Band 6) for most regions, except Cha I, Lupus, and Upper Sco which were observed at 338 GHz (Band 7). The choice of $\beta$ has no effect on the value of $\kappa_{\nu}$ for the surveys done in Band 6. For the surveys done in Band 7, using a $\beta$ of 0.4 yields a value of $\kappa_{\nu}$ that is $\sim$ 0.8 times the value obtained using a $\beta$ of 1, so the effect is minimal. Following \citet{manara22}, we adopt a $\beta$ of 1 for this analysis. We calculated two sets of disk masses, one using each dust temperature assumption. The left panel of Figure \ref{fig:massrecalc} shows the recalculated values using the $T_d$ scaling in Equation \ref{eq:td}; the right panel of Figure \ref{fig:massrecalc} shows recalculated values using a constant $T_d$ of 20 K.

Generally, as shown in Figure \ref{fig:massrecalc}, the masses inferred from SED modeling are greater than the masses obtained via Equation \ref{eq:mass}. A growing number of studies report that Equation 6 underestimates the dust disk mass. In their study of disks in Taurus using radiative transfer models, \citet{ballering19} found that Equation \ref{eq:mass} underestimates disk masses by a factor of $\sim$1-5.  Similarly, using ANN, \citet{ribas20} found disk masses greater than those reported in \citet{andrews13} (calculated from Equation \ref{eq:mass}) by a factor of $\sim$3. \citet{macias21} also found 3-5 times higher masses in TW Hya when comparing previous mm-flux-based mass estimates with a dust mass obtained from a radially resolved modelling of multi-wavelength ALMA observations that accounted for the radial variations in the optical depth of the disk. Most recently, \citet{liu22} investigated the effects of disk structure and dust properties on disk mass estimates through a radiative transfer parameter study; they find that mm-flux-based mass estimates typically underestimate disk masses by a factor of a few.

For the majority of disks in our study, our model-derived masses are 1.5-5 times larger than those obtained using Equation \ref{eq:mass}, consistent with \citet{ballering19}, \citet{ribas20}, \citet{macias21}, and \citet{liu22}.  Our result confirms the findings of these works: the assumptions required by Equation \ref{eq:mass} can lead to an underestimate of the disk mass.  Since radiative transfer models, and therefore ANN, do not rely on these assumptions, our disk masses do not suffer the same underestimation.

Given the growing list of ALMA surveys of T Tauri disks, a method to determine disk masses from millimeter photometry, such as Equation \ref{eq:mass}, is a useful tool. We present here a correction to Equation \ref{eq:mass}, based on the relationships in Figure \ref{fig:massrecalc}. In each panel of Figure \ref{fig:massrecalc}, we used the \texttt{LinearRegression} model in the \texttt{scikit-learn} Python package to fit a line to the relationship between disk masses from our SED models and disk masses calculated using Equation \ref{eq:mass}. We repeated the linear fitting 500 times; the best-fit lines and their uncertainties reported here represent the mean and standard deviation of the coefficients from the 500 linear fits. Given a disk mass $M$ in Earth masses calculated from millimeter photometry using Equation \ref{eq:mass}, the slopes and intercepts of the best-fit lines in Figure \ref{fig:massrecalc} can be used to obtain a new $M_{corrected}$ in Earth masses:

\begin{equation}\label{mcorr1}
    log_{10}(M_{corrected}) = 1.09(\pm 0.02) * log_{10}(M) + 0.19(\pm 0.02)
\end{equation}

\noindent if the disk temperature was scaled by the stellar luminosity, or

\begin{equation}\label{mcorr2}
    log_{10}(M_{corrected}) = 0.97(\pm 0.02) * log_{10}(M) + 0.29(\pm 0.01)
\end{equation}

\noindent if the disk temperature was assumed to be a constant 20 K.

While nearly all of our disk masses from SED modeling are larger than the recalculated literature disk masses, the discrepancy is not uniform when disk temperatures are scaled by stellar luminosity. As shown by the equations presented above, we find that disks with greater dust masses tend to deviate more from the recalculated literature values than disks with lower masses in the case where disk temperatures are scaled by the stellar luminosity. More luminous stars can supply more heat to the dust in the disk, increasing $T_d$. Assuming a constant dust temperature may result in an underestimate of $T_d$ for more massive stars, which, following Equation \ref{eq:mass}, would in principle result in an overestimate of $M_{dust}$. This has the result of flattening the linear relationship shown in Figure \ref{fig:massrecalc}(b), since the more massive disks tend to be around more massive stars \citep{andrews13, manara22}. If we account for this effect by scaling $T_d$ with stellar luminosity, we observe the trend of greater discrepancy at higher masses (see Figure \ref{fig:massrecalc}(a)). While it might seem that using the apparently incorrect fixed temperature can provide better results for more massive disks, we note that the trend is reversed for low mass disks: using a $T_d$ that scales with stellar luminosity provides a much better agreement with our SED-based masses. Regardless of the relationship between stellar luminosity and disk temperature, we find that Equation \ref{eq:mass} underestimates disk mass compared to our SED-derived masses, regardless of the relationship between stellar luminosity and disk temperature.

Another potential source of the discrepancy is the value of the reference dust opacity, $\kappa_{230\ GHz}$. As mentioned above, masses are typically calculated using a reference $\kappa_{230\ GHz}$ equal to 2.3 cm$^2$ g$^{-1}$, regardless of the size of the dust grains in the disk. The DIAD models compute different dust opacities depending on the size of the dust grains in the disk midplane, $a_{max,\ midplane}$. These dust opacities have a value of $\sim$2.3 cm$^2$ g$^{-1}$ for 400 \mum{} grains, $\sim$1.9 cm$^2$ g$^{-1}$ for a grain size of 1 mm, and a minimum of 0.85 cm$^2$ g$^{-1}$ for 1 cm grains. In most disks in our sample, $a_{max,\ midplane}$ is not well-constrained, and in fact the posteriors are usually flat from 100 microns to 1 cm. The difference in opacity hence enlarges (and mostly sets) our error bars in dust mass, making sure that the uncertainties in dust opacity are properly accounted for. Even in objects where $a_{max,\ midplane}$ might be relatively well constrained, the different opacities would only be able to explain mass differences a factor of 1.5-2.

\begin{figure*}
    \gridline{\fig{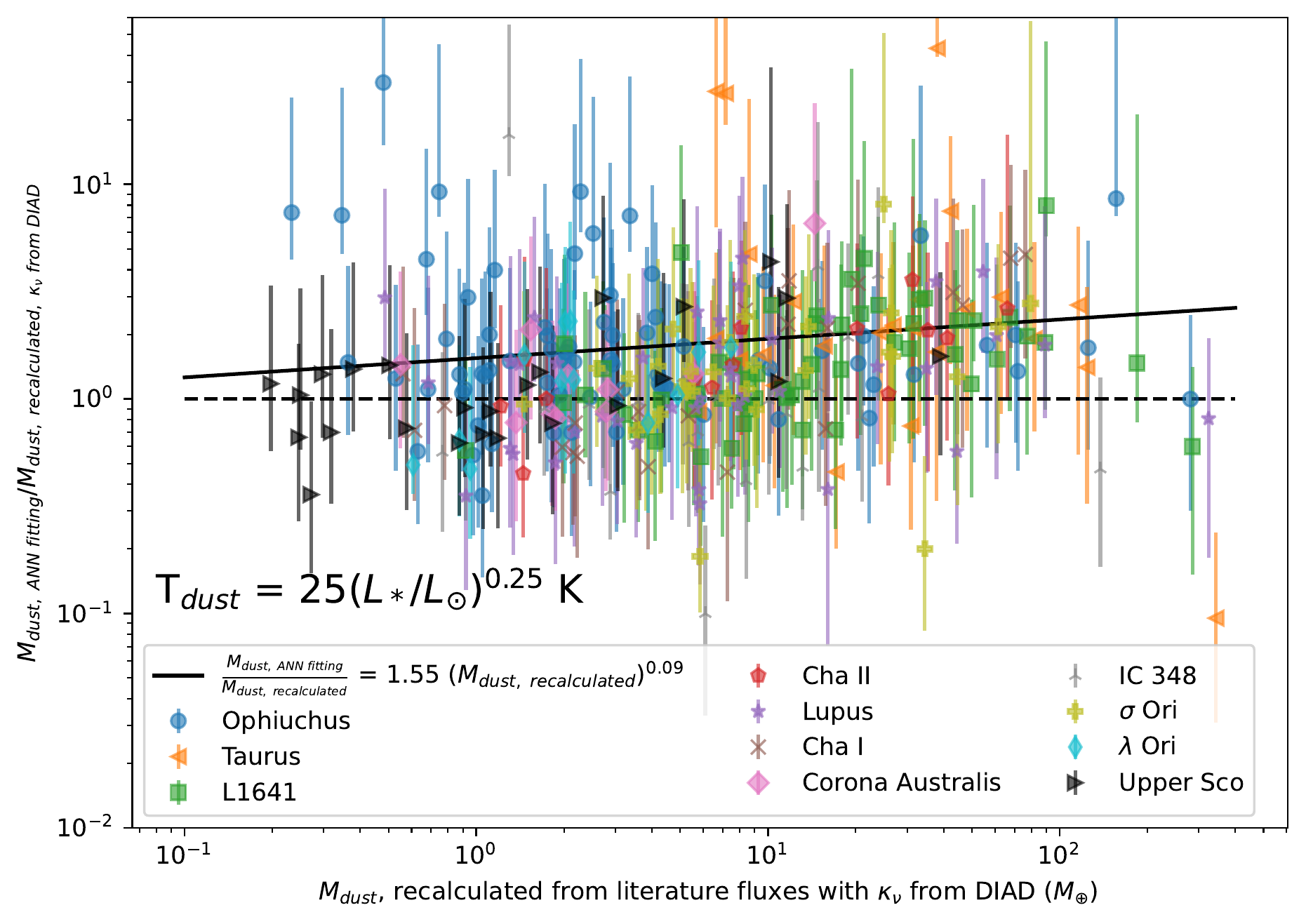}{0.5\textwidth}{(a)}
            \fig{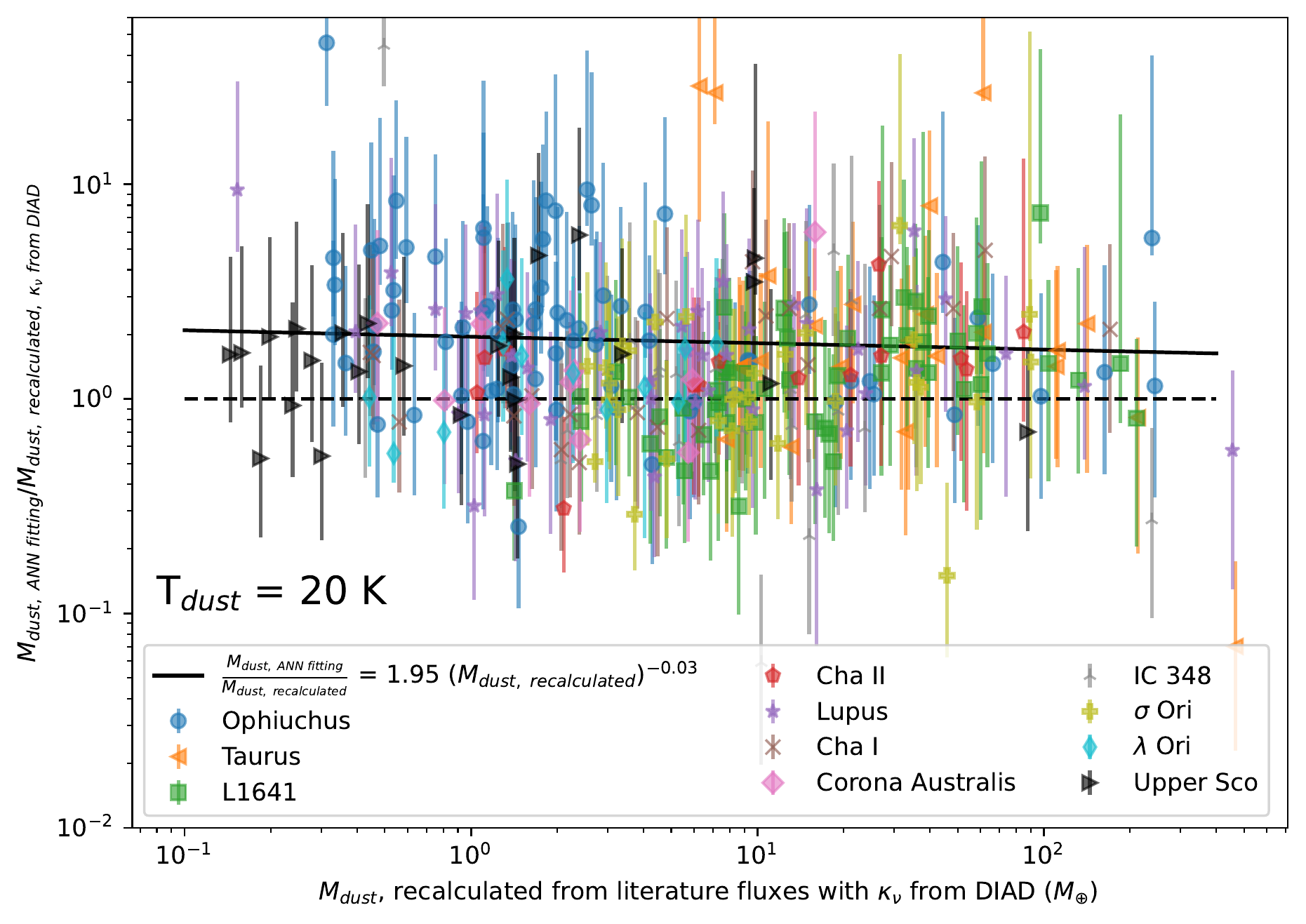}{0.5\textwidth}{(b)}}
    \caption{Comparison between our disk masses from SED modeling and disk masses calculated from single millimeter-wavelength fluxes, varying $\kappa_{230\ GHz}$ according to the median $a_{max,\ midplane}$ value of each object. The y-axis shows the ratio of our SED-derived masses to the flux-derived masses. The dashed black line shows a one-to-one relationship between the two masses, and the solid black lines show the linear fits from Figure \ref{fig:massrecalc}. As in Figure \ref{fig:massrecalc}, dust temperatures are scaled according to the stellar luminosity in Panel (a) and held at a constant value in Panel (b). In both cases, even when correcting for a more appropriate $\kappa_{230\ GHz}$ value, the dust masses we derive from SED fitting are still generally larger than those calculated from Equation \ref{eq:mass}.}
    \label{fig:kappacomp}
\end{figure*}

To probe the effect of the value of $\kappa_{230\ GHz}$ on dust mass, we again recalculated masses using Equation \ref{eq:mass}. This time, we used the appropriate $\kappa_{230\ GHz}$ for each object as calculated by DIAD for its median $a_{max,\ midplane}$. All other assumptions were kept the same as described above and in \citet{manara22}. Figure \ref{fig:kappacomp} shows the ratio of our SED-derived dust masses to these new recalculated dust masses plotted versus the new recalculated dust masses. As can be seen in the Figure, our SED-derived dust masses are still a few times greater than those calculated from Equation \ref{eq:mass}, even when accounting for different $\kappa_{230\ GHz}$. We note that uncertainties in dust composition and structure could affect the value of both \bm{$\kappa_{230\ GHz}$} and \bm{$\beta$} by as much as an order of magnitude \citep{testi14, miotello22}, but these would have the same effect both on our dust masses and the millimeter-flux ones. While these uncertainties could alter the dust masses, the differences in $\kappa_{230\ GHz}$ alone cannot explain the observed discrepancy between our dust masses and those derived from millimeter fluxes. Differences in $\kappa_{230\ GHz}$ cannot alone explain the observed discrepancy between our dust masses and those derived from millimeter fluxes.

An alternative explanation for the observed trend is related to the assumption made in Equation \ref{eq:mass} that the disk is optically thin at the reference frequency.  More massive disks are expected to be optically thicker at a given wavelength due to the greater amount of absorbing material present; if these disks are optically thick, Equation \ref{eq:mass} will underestimate the disk mass since the observed flux is not sensitive to the whole mass in the disk. Modeling by \citet{mohanty13}, for example, showed that the 850 micron flux density becomes independent of disk mass in the optically thick limit. \citet{ballering19}, \citet{ribas20}, and \citet{liu22} all cite optical depth as a way to explain the discrepancy between disk mass estimates from (sub)mm fluxes and those from disk SED modeling. As mentioned above in Section \ref{ann}, the SED models calculate the full disk structure, which includes the optical depth. Therefore, even when the observed millimeter emission is partially optically thick, the simulated millimeter emission takes this into account, and thus the total dust mass can be constrained.

\begin{figure*}
    \gridline{\fig{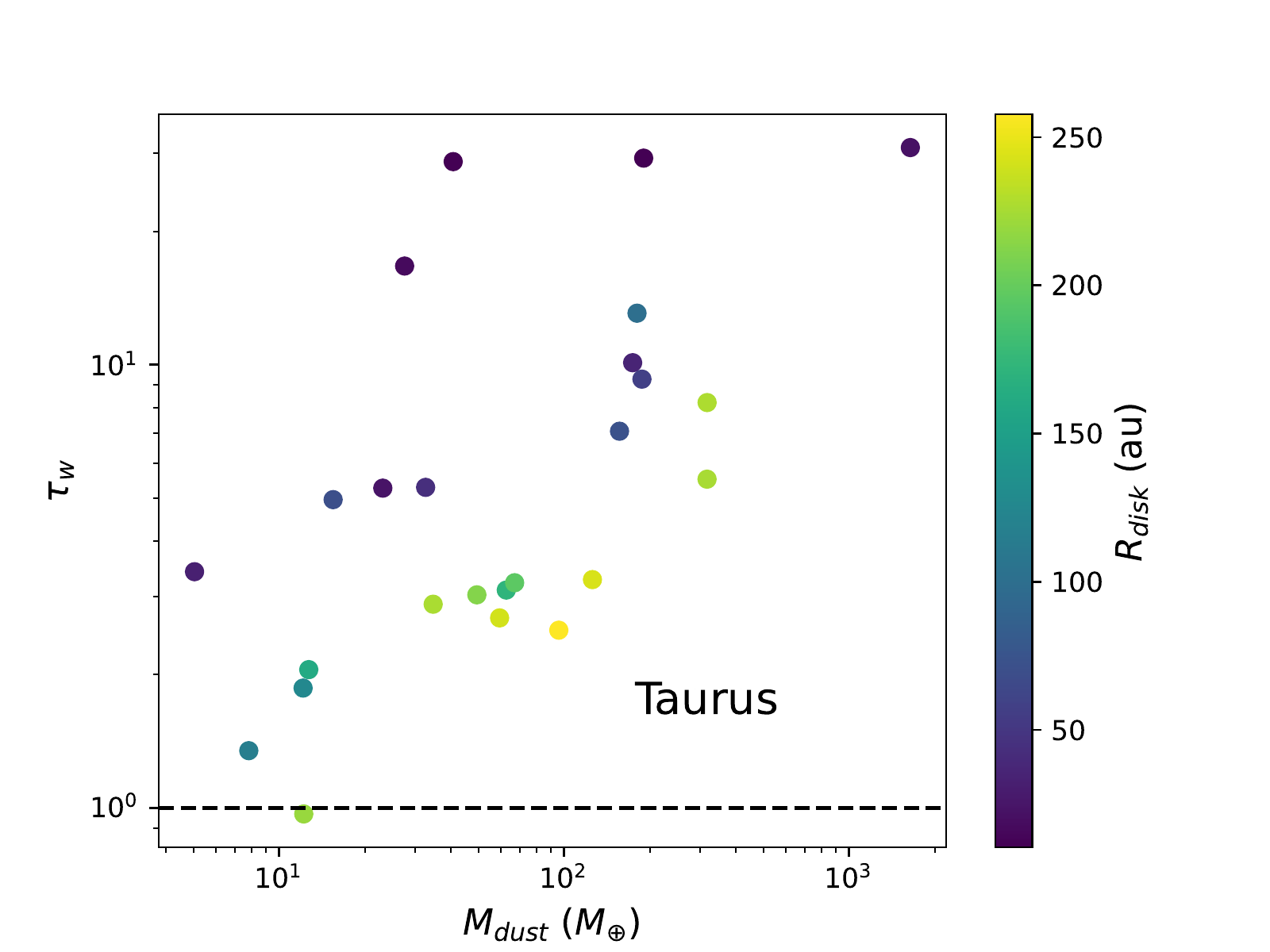}{0.5\textwidth}{(a)}
            \fig{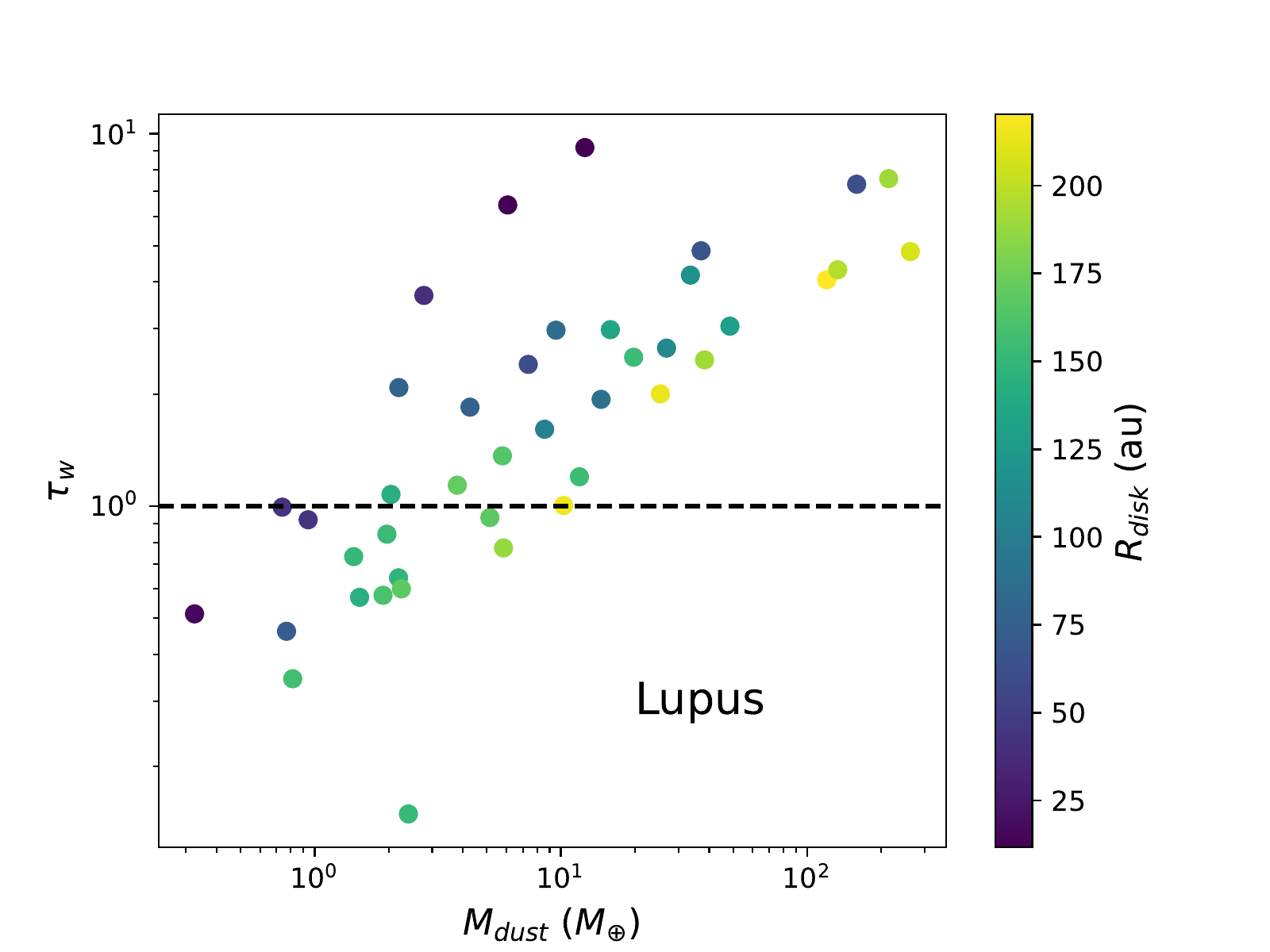}{0.5\textwidth}{(b)}}
    \vspace{-0.5cm}
    \gridline{\fig{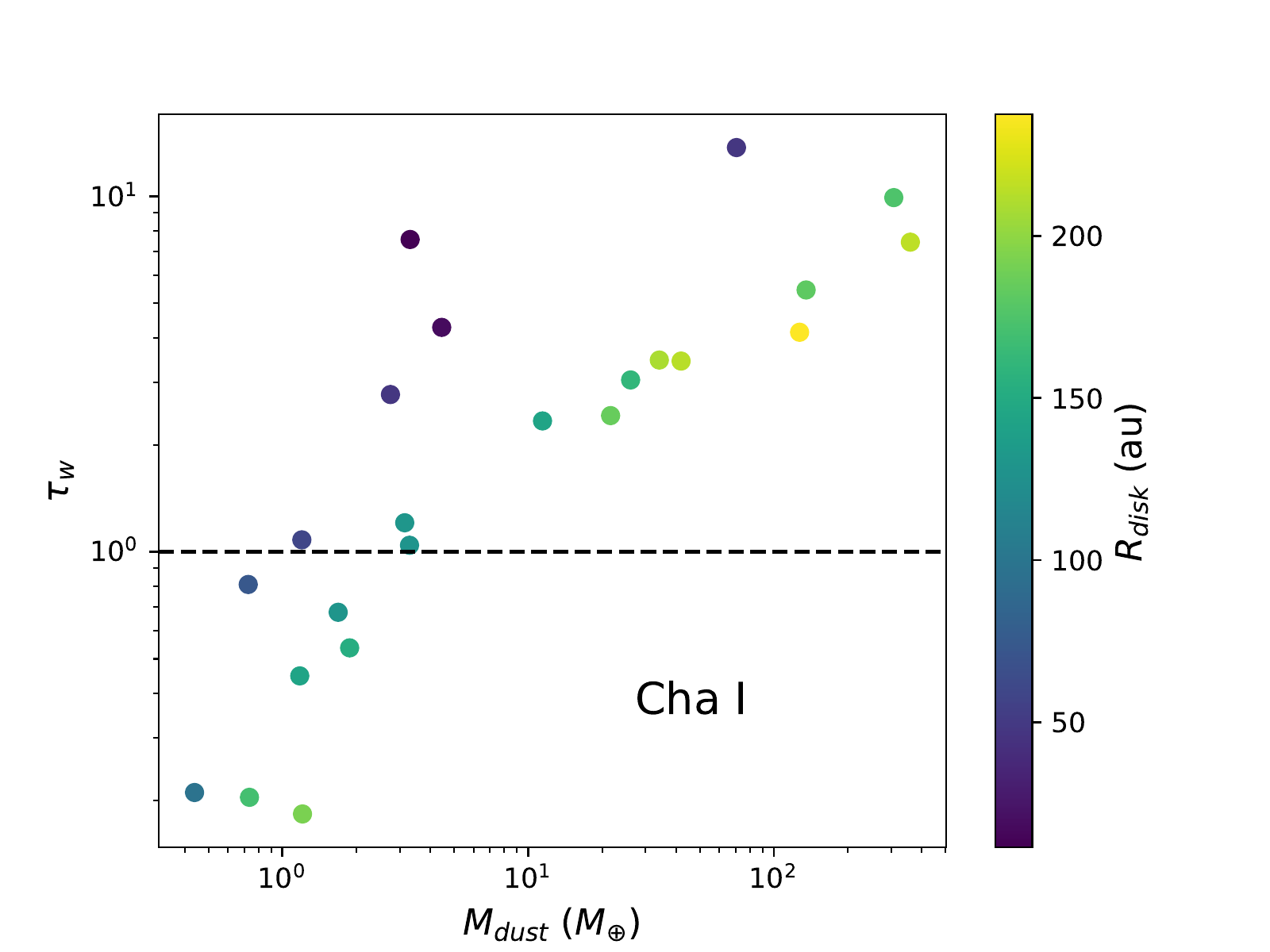}{0.5\textwidth}{(c)}
            \fig{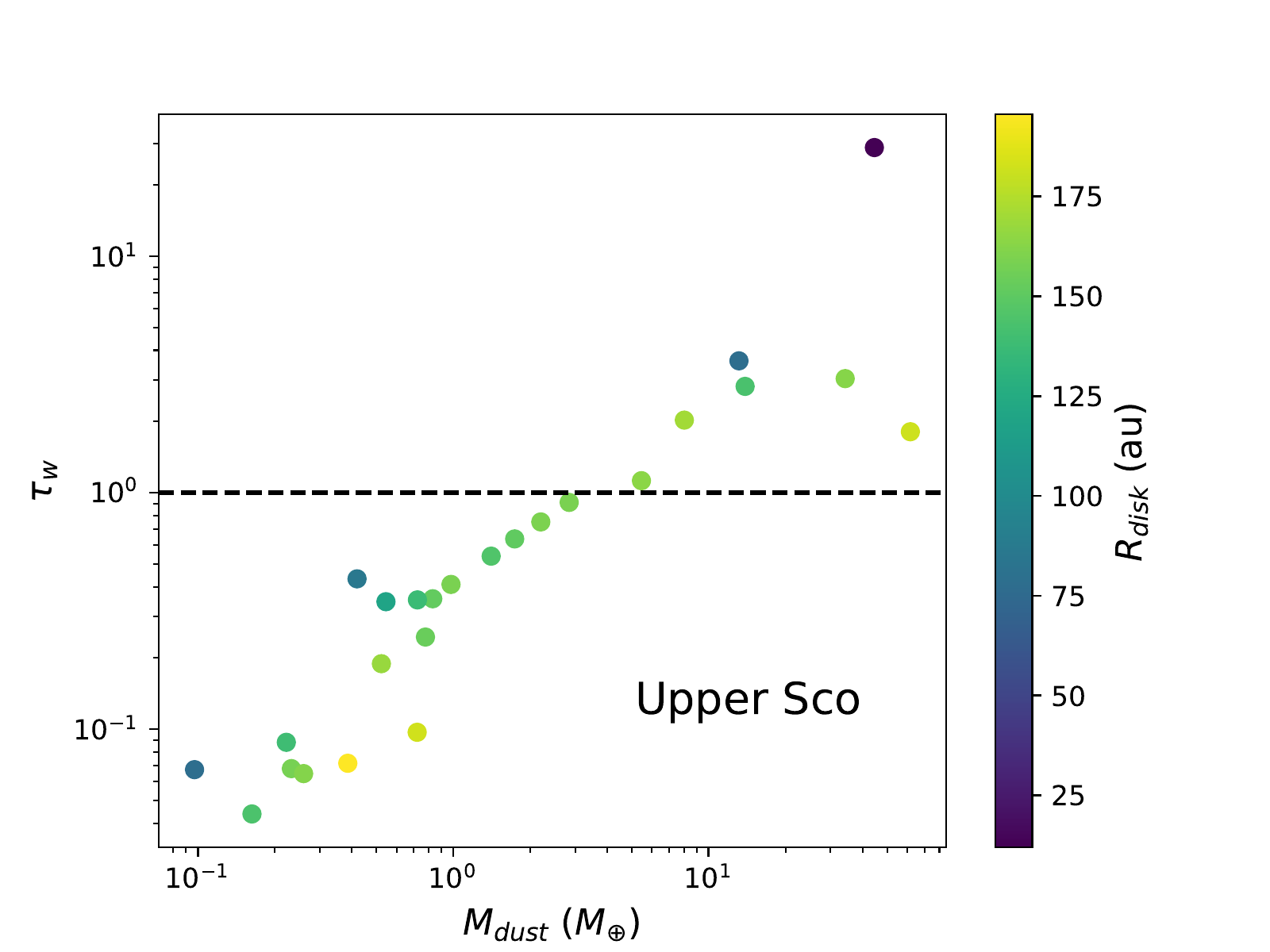}{0.5\textwidth}{(d)}}
    \caption{Flux-weighted mean optical depth of each disk plotted versus disk mass for four star-forming regions: (a) Taurus, (b) Lupus, (c) Cha I, and (d) Upper Sco.  Optical depths are flux-weighted averages of the 1.3 mm optical depth at each radius in the disk as calculated by the DIAD models.  Disk masses are the median values from our models for each object. The horizontal dashed line denotes an optical depth of 1. Colors represent the disk outer radii. Some disks in each region have optical depth greater than 1, especially higher mass disks.}
    \label{fig:optdepth}
\end{figure*}

To test this explanation, we ran DIAD models for a subset of our disk sample in order to calculate optical depths. Using the median values for each model parameter from our fits, we used DIAD to calculate the structure of the disk and optical depth along the line of sight as a function of radius at 225 GHz (1.3 mm). We then computed a flux-weighted mean optical depth using the flux at each radius as the weight for the optical depth at that radius. In this way, we obtained a single average optical depth of the disk that takes into account the regions of the disk where most of the emission comes from. These flux-weighted mean optical depths are plotted versus dust mass in Figure \ref{fig:optdepth} for each disk in Taurus, Lupus, Cha I and Upper Sco. (We selected these regions since they are the four regions studied by \citet{pascucci16}; we compare our $M_{dust}$ -- $M_*$ relationships to theirs in the following section.) In each region, optical depth generally increases with disk mass. Many disks in each region have a flux-weighted mean optical depth close to or greater than one, indicating that the optically thin assumption is not valid for these disks. Thus, disk masses calculated from Equation \ref{eq:mass} are systematically underestimated; disk masses obtained with DIAD and ANN$_{diskmass}$, which do not assume the disks are optically thin, are more reliable for determining disk masses.

We also show the disk size for each disk in Figure \ref{fig:optdepth} by coloring each point according to its median outer radius from the MCMC fitting. These outer radii are larger than the FWHMs reported for resolved disks in \citet{ansdell16}, \citet{pascucci16}, and \citet{barenfeld16}, though larger FWHMs generally correspond to larger outer radii in our sample. Disks with lower masses but high optical depths have smaller radii, indicating that they are compact and dense, while disks with higher masses but lower optical depths have larger radii and are comparatively less dense.

\subsubsection{Impact on Planet Formation Models}
The higher masses resulting from our SED models can help alleviate the tension between the observed masses of exoplanetary systems and protoplanetary disks.  Previously, \citet{manara18} reported that the dust disks in Lupus and Cha I are a factor of 3--5 too small to account for the typical mass of planetary cores. A more recent study by \citet{mulders21} reports more comparable masses between disks and exoplanetary systems, but requires nearly 100\% efficiency in the planet formation process to explain the observed masses. The masses we report here are typically a factor of 1.5--5 greater than the recalculated literature values. The assumption that disks are optically thin can therefore result in a significant underestimation of their masses. Our disk masses ease the requirement for extremely high planet formation efficiency. We note, however, that increasing the available disk mass by a factor of 1.5-5 would imply planet formation efficiencies between $\sim$20\% and $\sim$60\%, to achieve the same exoplanet disk masses as reported by \citet{mulders21}. Given that planet formation models do not still have a robust estimate of the planet formation efficiency, it is not fully clear if the optical depth alone could solve the ``missing'' mass problem or if alternative scenarios are still required (e.g., planetesimals already being formed, late infall of material onto the disk).

\subsection{Disk Mass Trends with Host Mass}\label{masstrend}

\begin{figure*}
    \gridline{\fig{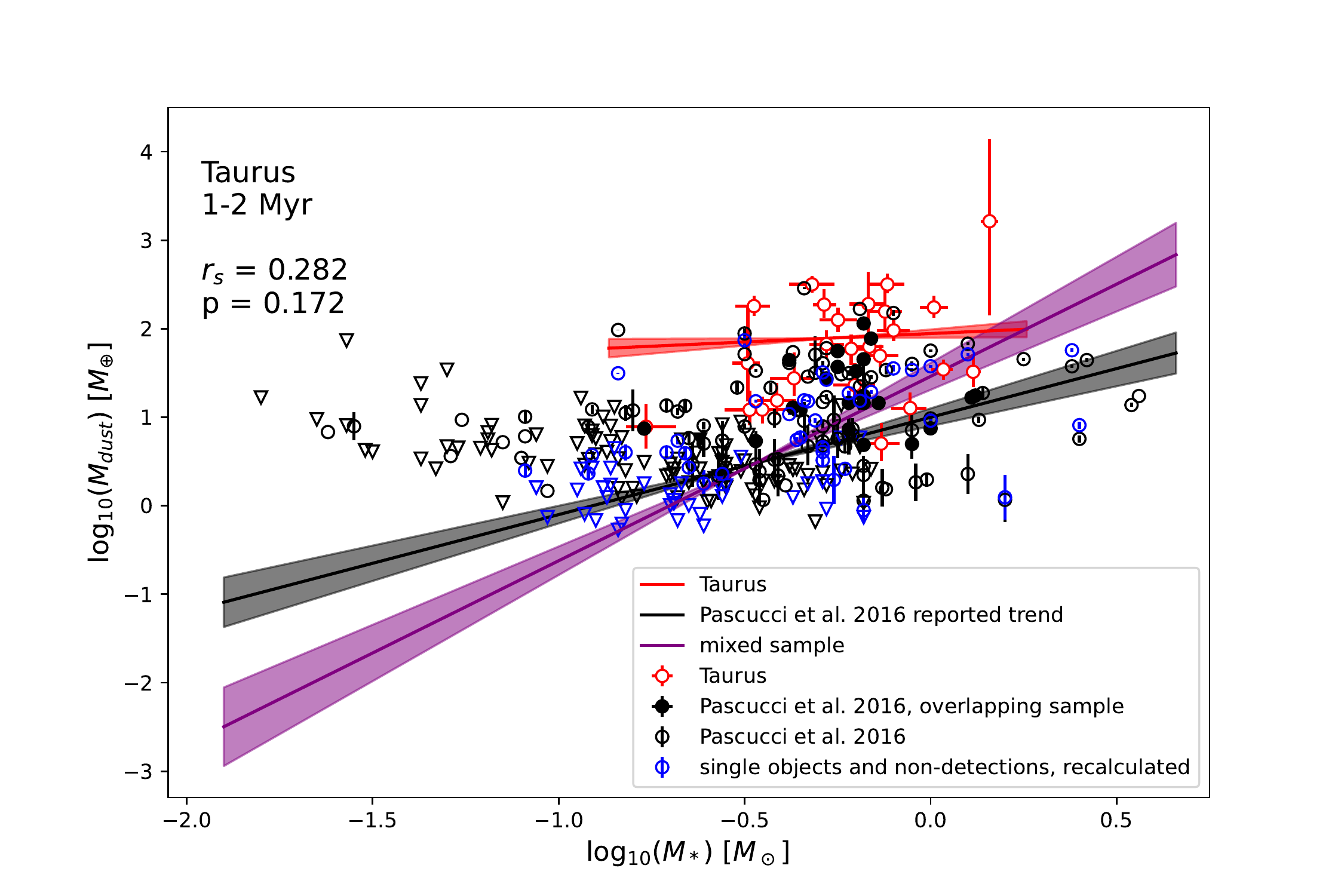}{0.53\textwidth}{(a)}
            \hspace{-0.5cm}
            \fig{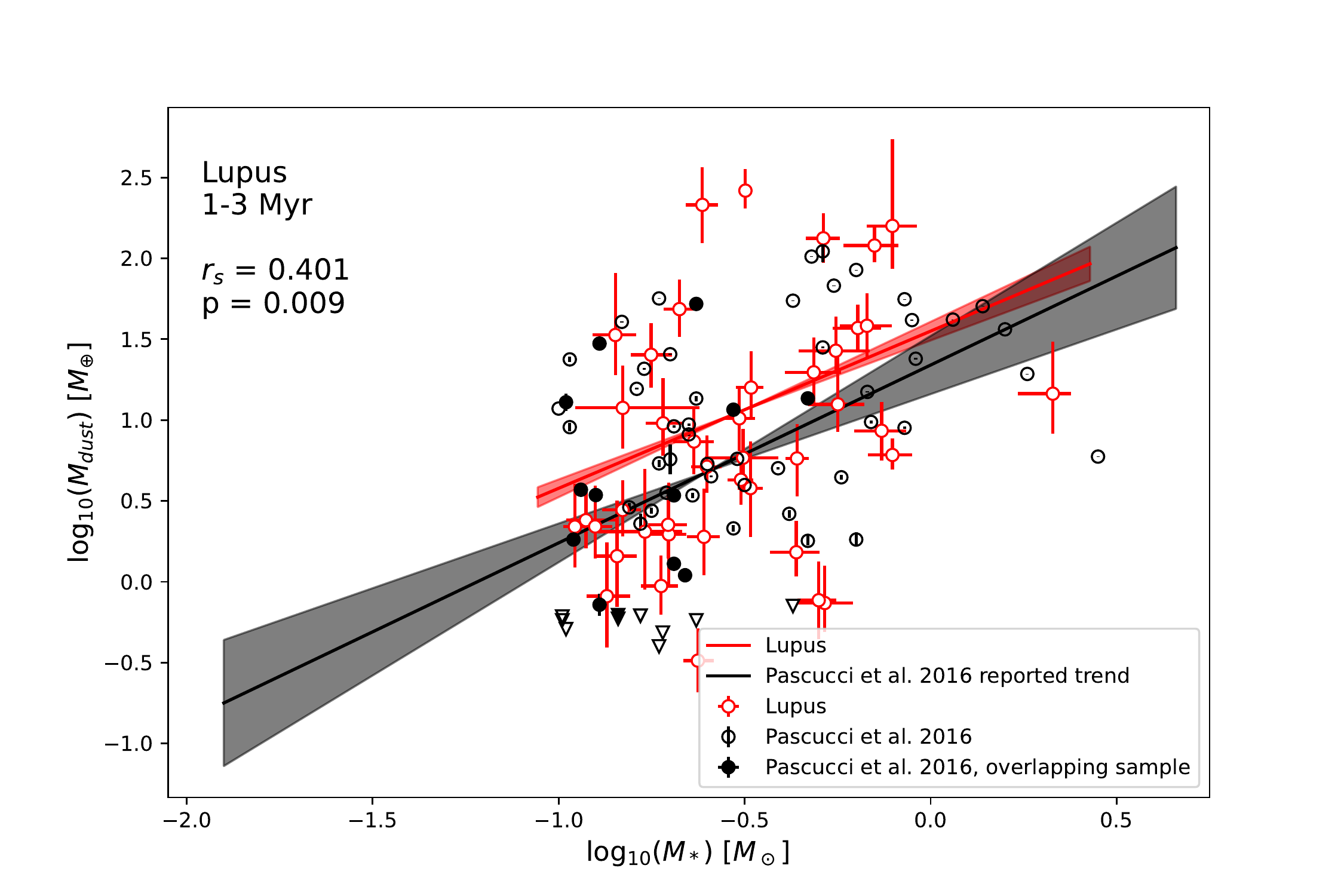}{0.53\textwidth}{(b)}}
    \vspace{-0.5cm}
    \gridline{\fig{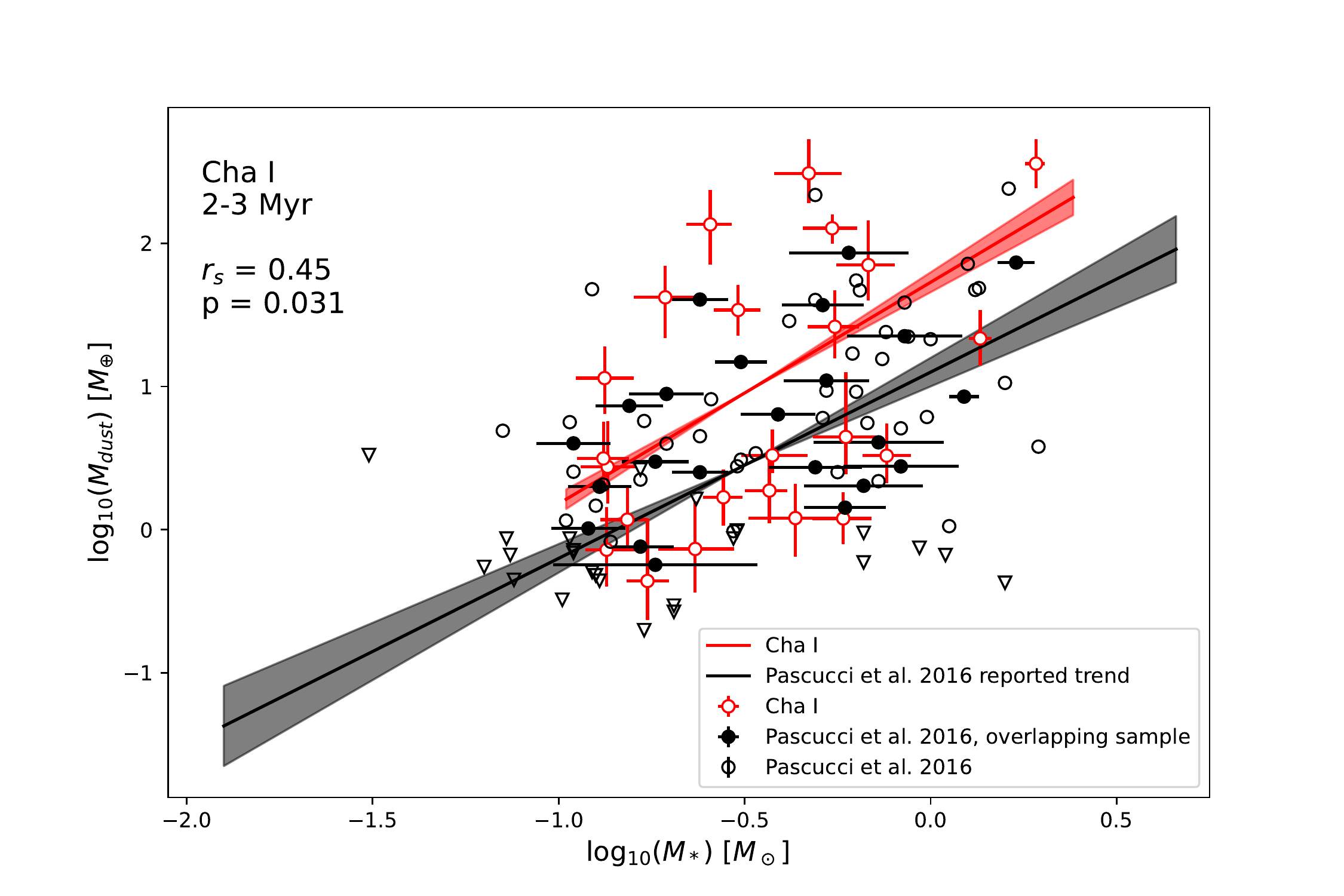}{0.53\textwidth}{(c)}
            \hspace{-0.5cm}
            \fig{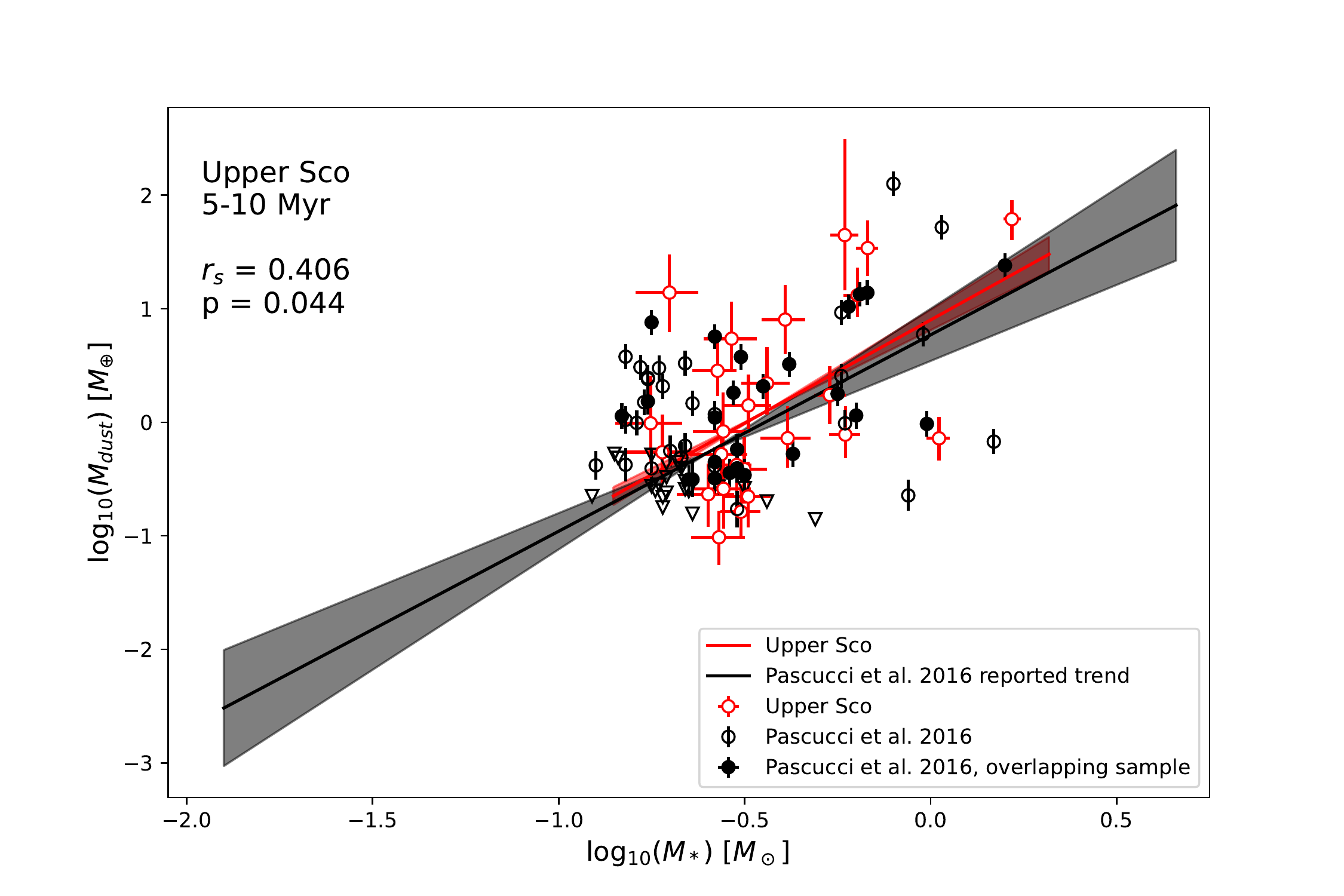}{0.53\textwidth}{(d)}}
    \caption{The relationship between disk dust masses and host star masses for four star-forming regions: (a) Taurus, (b) Lupus, (c), Cha I, and (d) Upper Sco. Red points are median $M_{dust}$ and $M_*$ values from the MCMC output and black points are $M_{dust}$ and $M_*$ values from \citep{pascucci16}. Linear fits for these two groups are shown in their matching colors, with a 1$\sigma$ uncertainty shown by the shaded region. The values listed in the upper left corner of each panel are the Spearman correlation coefficient and p value for our modeled $M_{dust}$ and $M_*$ values. Taurus has a lower fraction of modeled objects than other regions, so we also include a fit to a mixed sample (purple line): our modeled objects plus single objects and non-detections not included in our sample, with masses recalculated from their mm fluxes (blue symbols, see Section \ref{masscomp}).}
    \label{fig:mdiskmstar-pasc}
\end{figure*}

Disk masses have been shown to scale with the masses of their host stars in individual star forming regions \citep[e.g.,][]{pascucci16, testi22}, though the trend is murkier when multiple regions are considered together \citep{manara22}.  In Figure \ref{fig:mdiskmstar-pasc} we plot disk dust mass versus host mass for the four star-forming regions studied by \citet{pascucci16}. Following \citet{pascucci16}, we use the Bayesian linear fitting procedure developed by \citet{kelly07}, which allows for measurement uncertainties as well as upper limits, to fit the recalculated dust mass relationship. Our MCMC fitting process results in asymmetric uncertainties, which are not supported by the \citet{kelly07} fitting procedure, so we instead use Python Abstract Interfaces for Data Analysis (PAIDA)\footnote{paida.sourceforge.net} to obtain a linear fit to the modeled dust masses that incorporates the uncertainties.

For Lupus, Cha I, and Upper Sco, we find $M_{dust}$-$M_*$ relationships that are consistent with those reported in \citet{pascucci16}; i.e., the slopes and intercepts of the fits reported in Table \ref{tab:mdiskmstar} for each region are within 1-2$\sigma$ of the fits reported by \citet{pascucci16}. We generally find slightly larger intercepts than \citet{pascucci16}; this can be attributed to the result reported above, that we find larger disk masses than previously reported. The relationship we find for disks in Taurus is shallower than previously reported. This difference in slope may be due to the inherent bias in our sample introduced by the exclusion of upper limits. Since most the disks in Taurus were observed at lower sensitivity by the SMA, many of these objects only have upper limits for their disk masses; the linear fit for Taurus reported by \citet{pascucci16} incorporates these points, but our sample does not and is limited to higher mass objects.  

To mediate the effect of this difference in sensitivity, we also present an $M_{dust}$-$M_*$ relationship for a mixed sample of Taurus objects. This mixed sample includes our modeled objects, plus any detected objects not known to be in binary or multiple systems but excluded for other reasons, and single non-detections. We adopt host and disk masses from our SED models where possible; for the remaining objects, host masses were taken from \citet{pascucci16} and disk masses were recalculated from literature fluxes using our consistent method described in Section \ref{masscomp}, with a constant dust temperature of 20 K. The $M_{dust}$-$M_*$ relationship of this mixed sample, shown in Figure \ref{fig:mdiskmstar-pasc}(a) is more consistent with the relationship from \citet{pascucci16}, as well as the other three regions presented in Figure \ref{fig:mdiskmstar-pasc}.

We note that the correlation of the $M_{dust}$-$M_*$ relationship in each region is only moderate at best.  Spearman correlation coefficients ($r_s$) for our modeled $M_{dust}$ and $M_*$ values are given in the upper left corner of each panel in Figure \ref{fig:mdiskmstar-pasc} and range between $\sim$ 0.3 and $\sim$ 0.5. The $r_s$ value describes how well the relationship between $M_{dust}$ and $M_*$ is described by a monotonic function; $r_s$ values closer to 1 indicate stronger correlation. For each region, we also report the p value, which gives the probability that a random, uncorrelated sample would produce a similar correlation. For Taurus, the $r_s$ and p values given in Figure \ref{fig:mdiskmstar-pasc} are for our sample only; for the mixed sample, $r_s$ = 0.425 and p = 0.001. Though the $r_s$ values show only a moderate correlation, the low p values for each of these four regions indicates that the moderate correlation is statistically significant. \citet{pascucci16} also reported moderate correlations, with dispersions of $\sim$0.8 dex in the $M_{dust}$ -- $M_*$ relationship for each region.

\begin{figure*}
    \gridline{\fig{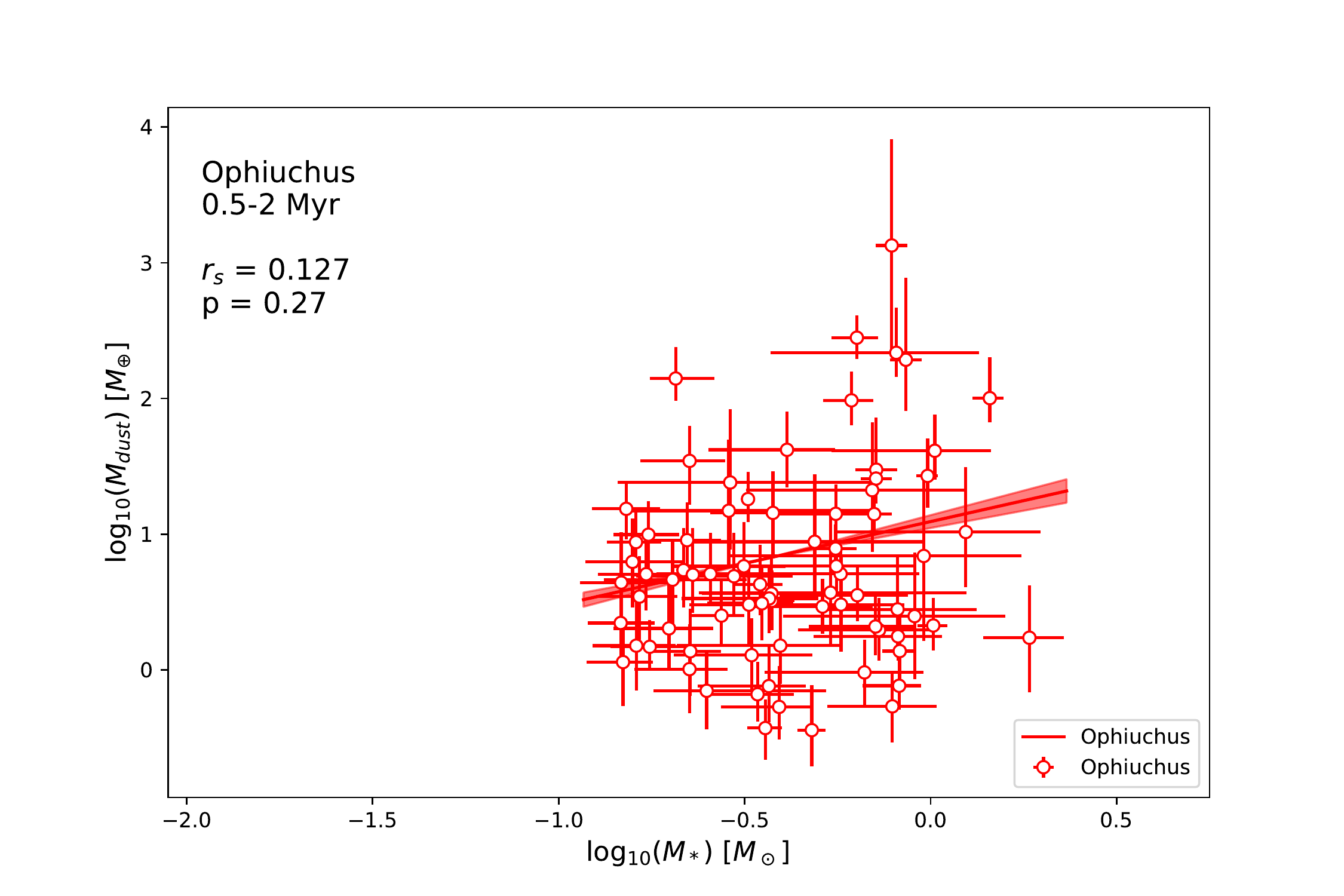}{0.35\textwidth}{(a)}
            \hspace{-0.7cm}
            \fig{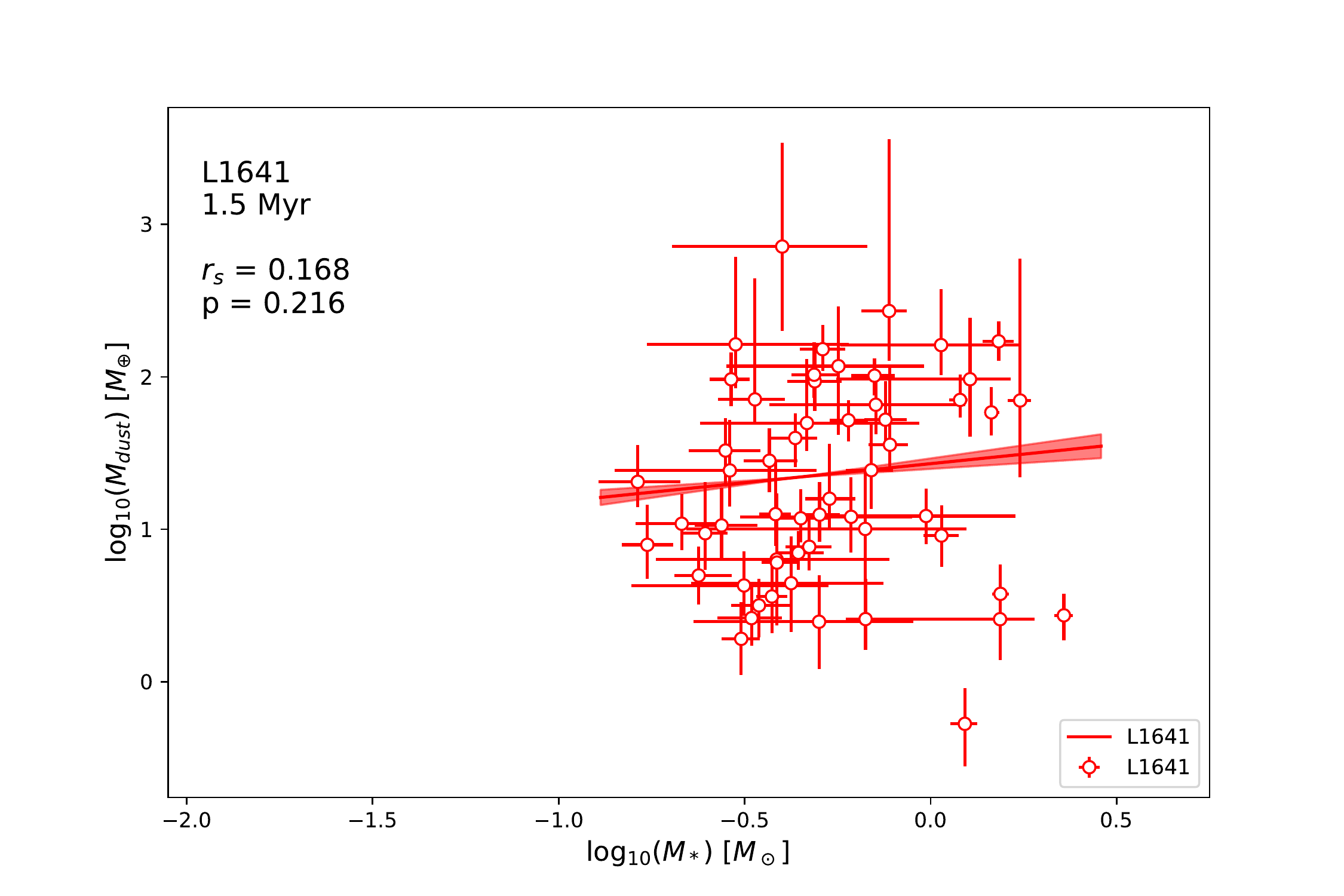}{0.35\textwidth}{(b)}
            \hspace{-0.7cm}
            \fig{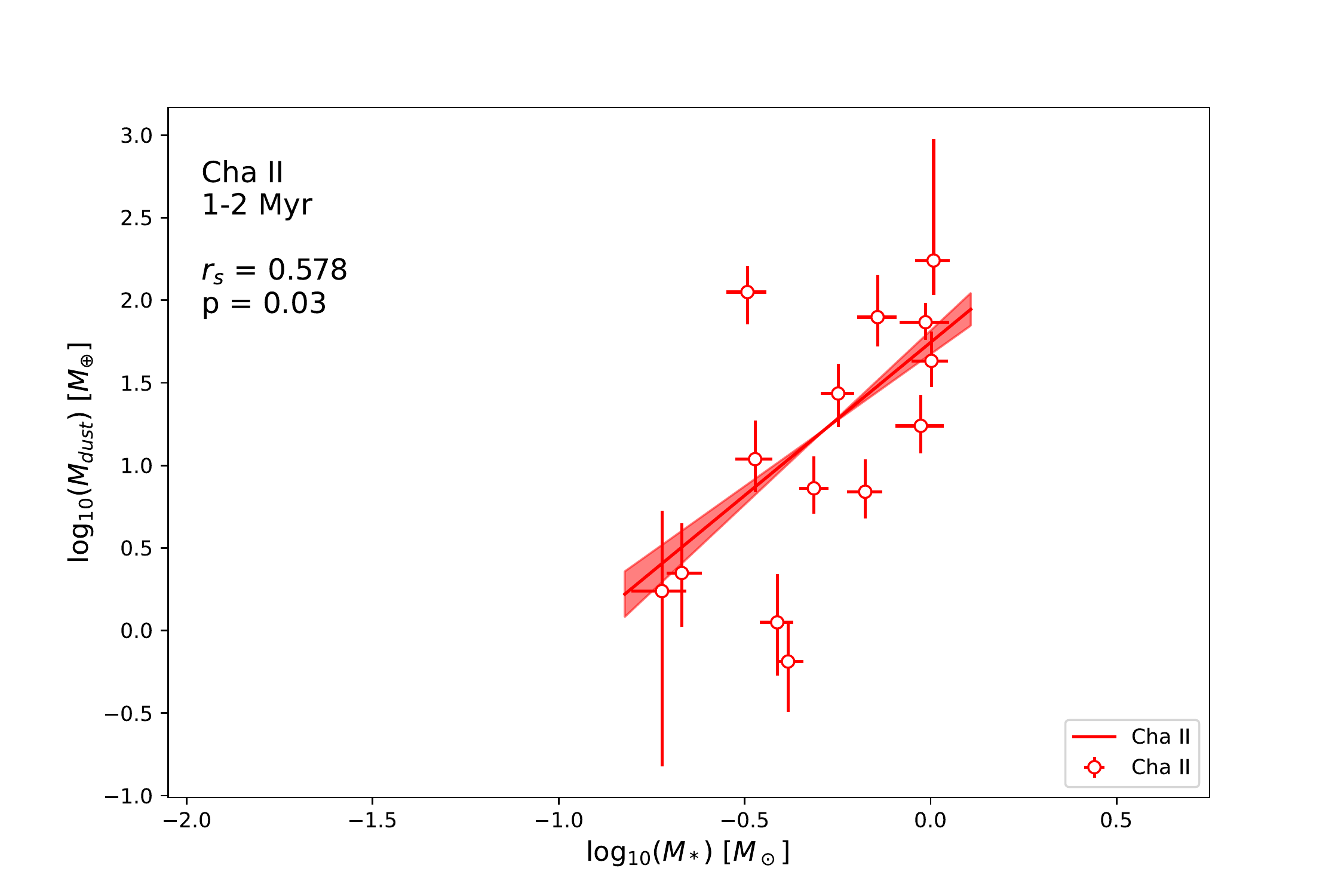}{0.35\textwidth}{(c)}}
    \vspace{-0.5cm}
    \gridline{\fig{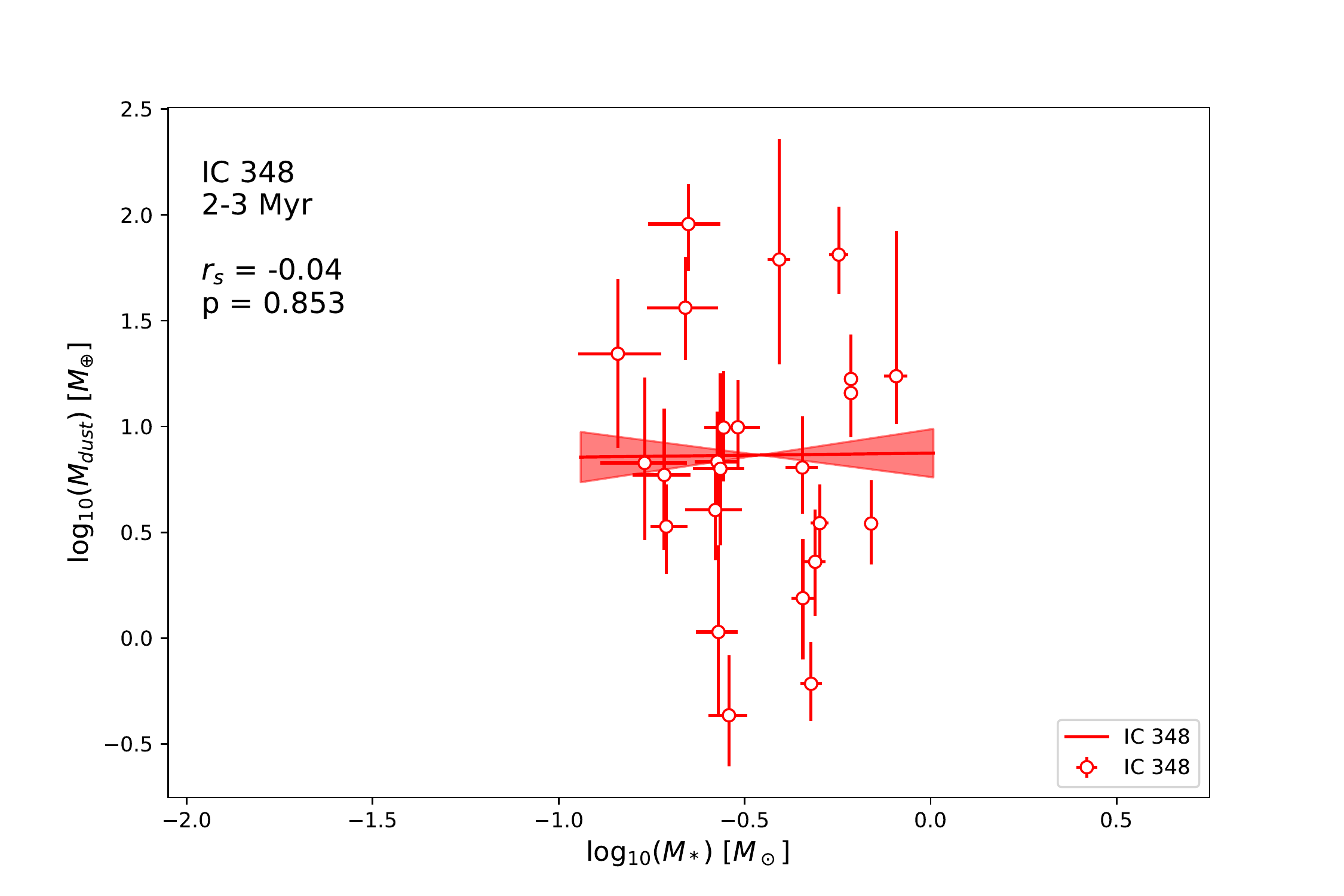}{0.35\textwidth}{(d)}
            \hspace{-0.7cm}
            \fig{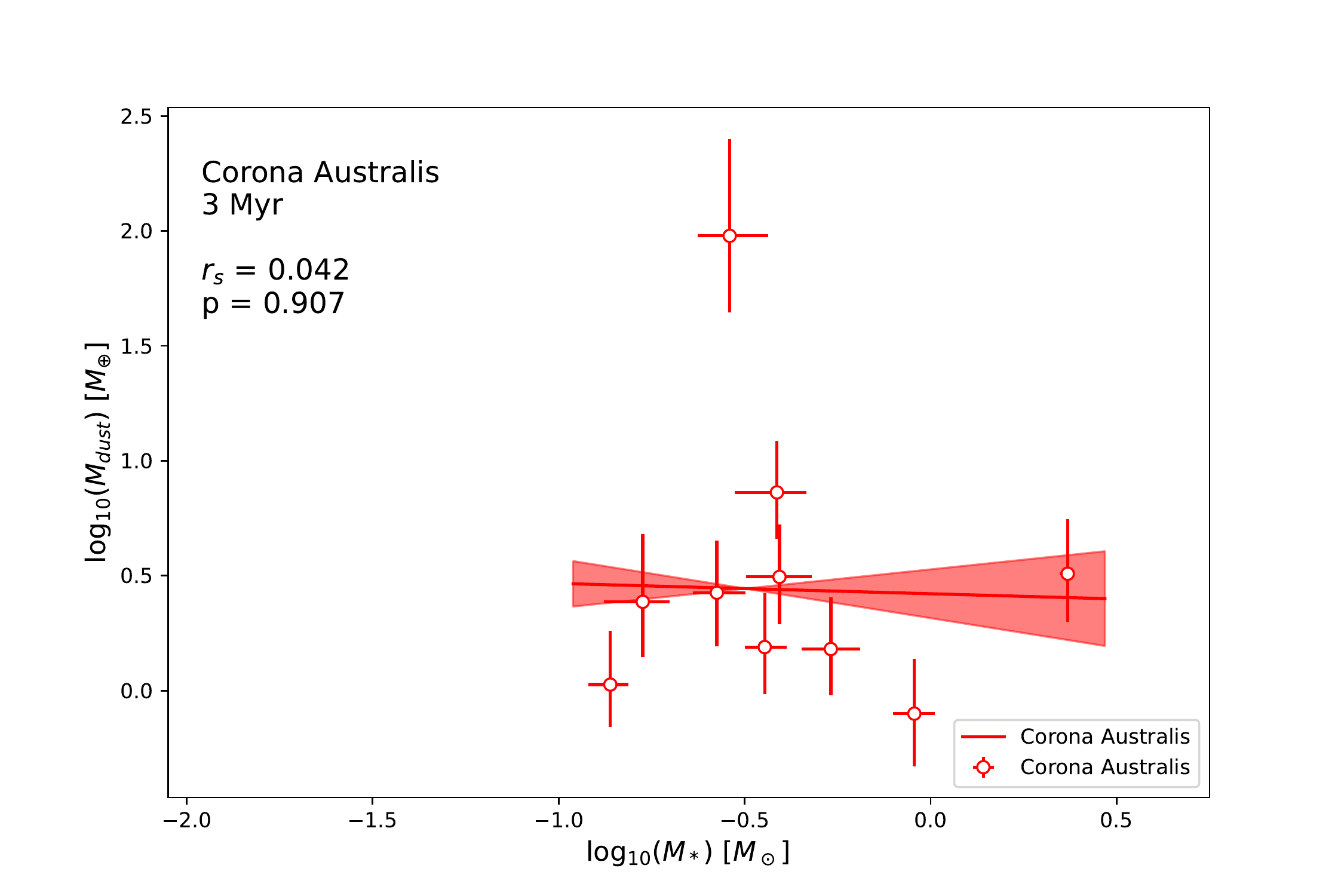}{0.35\textwidth}{(e)}
            \hspace{-0.7cm}
            \fig{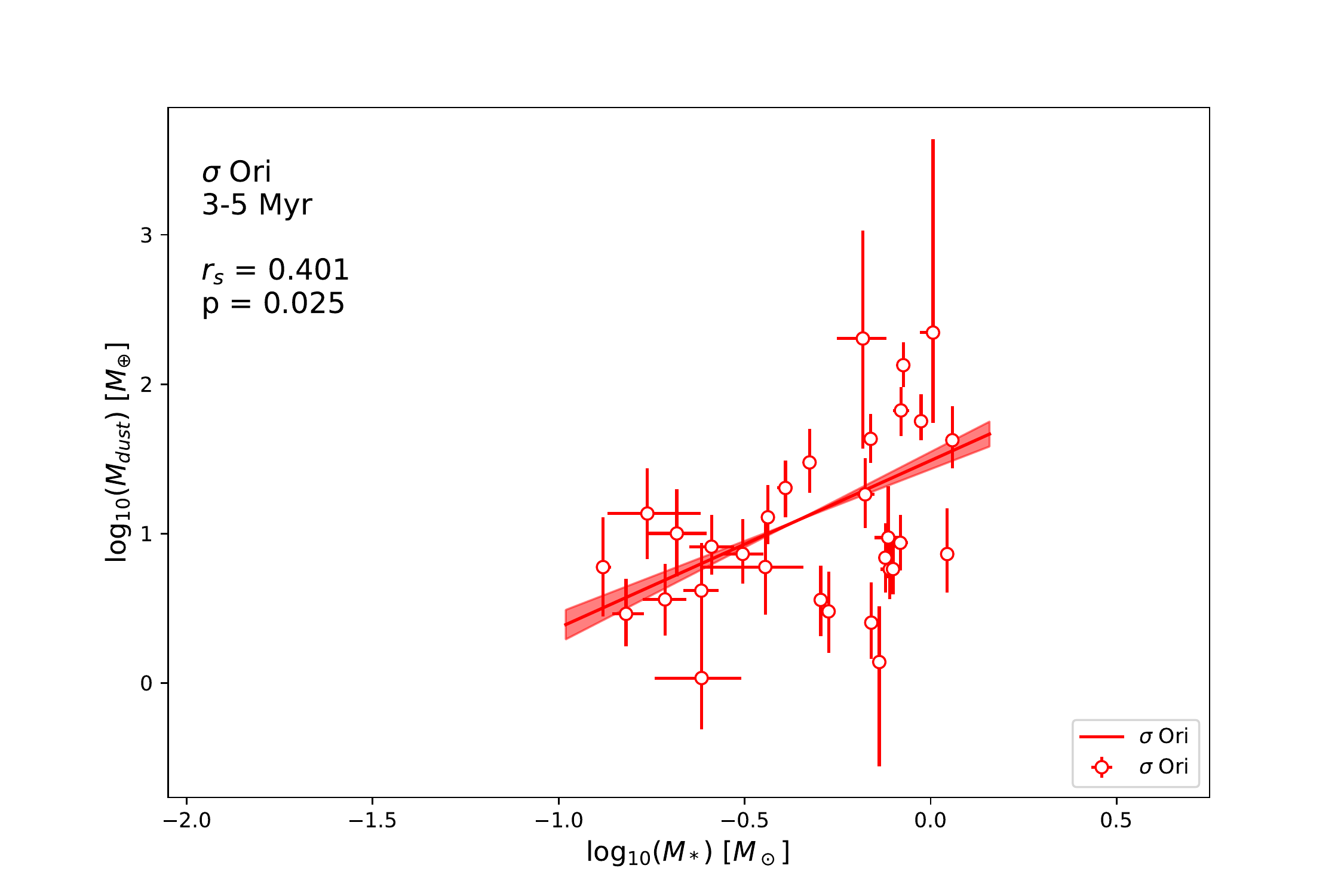}{0.35\textwidth}{(f)}}
    \vspace{-0.5cm}
    \gridline{\fig{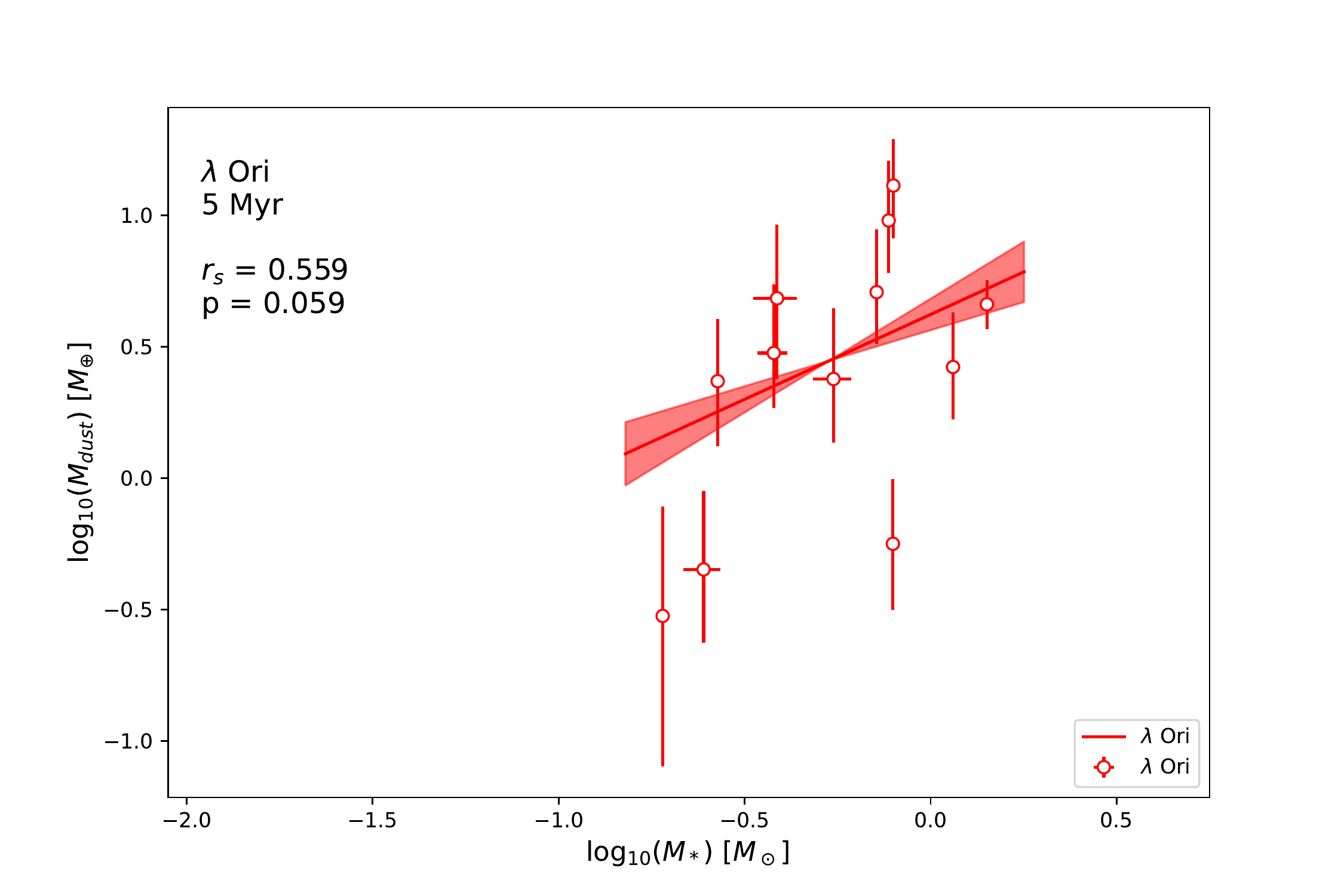}{0.35\textwidth}{(g)}}
    \caption{The relationship between disk dust masses and host star masses for the star-forming regions not studied by \citet{pascucci16}: (a) Ophiuchus, (b) L1641, (c), Cha II, (d) IC348, (e) Corona Australis, (f) $\sigma$ Ori, and (g) $\lambda$ Ori. Red points are median $M_{dust}$ and $M_*$ values from the MCMC output. Linear fits for each region are shown in red, with a 1$\sigma$ uncertainty shown by the shaded region. The values listed in the upper left corner of each panel are the Spearman correlation coefficient and p value for our modeled $M_{dust}$ and $M_*$ values.}
    \label{fig:mdiskmstar-other}
\end{figure*}

The $M_{dust}$-$M_*$ relationships for the remaining seven regions in our sample are shown with their $r_s$ and p values in Figure \ref{fig:mdiskmstar-other}; we report the slopes and intercepts of the relationships in Table \ref{tab:mdiskmstar}.  We note that for three regions (Cha II, Corona Australis, and $\lambda$ Ori) the sample sizes may be too small to accurately assess correlation. Of the four regions with substantial samples (Ophiuchus, L1641, IC 348, and $\sigma$ Ori), only one, $\sigma$ Ori, has a statistically significant trend (i.e., p $<$ 0.05). The $r_s$ value for this region is comparable to the $r_s$ values for the regions shown in Figure \ref{fig:mdiskmstar-pasc}, showing a moderate correlation between $M_{dust}$ and $M_*$. Ophiuchus, L1641, and IC 348 do not show any significant correlation. All eleven regions in this sample show a large dispersion in $M_{dust}$ for a given $M_*$. As noted in the review by \citet{manara22}, the fact that this dispersion is observed in all regions suggests it is due to inherent variation in disk properties, as opposed to environmental effects in a specific region.

\begin{deluxetable}{c c c c c}
\tablecaption{$M_{dust}$ -- $M_*$ Relationships\label{tab:mdiskmstar}}
\tablehead{
\colhead{Region} & \multicolumn{2}{c}{This work} & \multicolumn{2}{c}{\citet{pascucci16}}\\
\colhead{} & \colhead{$m$} & \colhead{$b$} & \colhead{$m$} & \colhead{$b$}}
\startdata
Ophiuchus TTS & 0.6$\pm$0.1 & 1.1$\pm$0.1 & ... & ...\\
Ophiuchus TTS and BDs & 0.9$\pm$0.1 & 1.2$\pm$0.1 & ... & ...\\	
Taurus TTS & 0.2$\pm$0.2 & 1.9$\pm$0.1 & 1.1$\pm$0.2 & 1.0$\pm$0.1\\
      & 2.1$\pm$0.3\tablenotemark{a} & 1.4$\pm$0.2\tablenotemark{a} & & \\
Taurus TTS and BDs & 1.6$\pm$0.1 & 2.1$\pm$0.1 & ... & ...\\
L1641 & 0.2$\pm$0.1 & 1.43$\pm$0.1 & ... & ...\\
Cha II & 1.9$\pm$0.2 & 1.7$\pm$0.1 & ... & ...\\	
Lupus TTS & 1.0$\pm$0.1 & 1.5$\pm$0.1 & 1.1$\pm$0.3 & 1.4$\pm$0.2\\
Lupus TTS and BDs & 1.1$\pm$0.1 & 1.3$\pm$0.1 & ... & ...\\
Cha I & 1.5$\pm$0.1 & 1.7$\pm$0.1 & 1.3$\pm$0.2 & 1.1$\pm$0.1\\
IC 348 & 0.0$\pm$0.2 & 0.9$\pm$0.1 & ... & ...\\
Corona Australis & 0.0$\pm$0.2 & 0.4$\pm$0.1 & ... & ...\\
$\sigma$ Ori & 1.1$\pm$0.2 & 1.5$\pm$0.1 & ... & ...\\
$\lambda$ Ori & 0.6$\pm$0.2 & 0.6$\pm$0.1 & ... & ...\\	
Upper Sco TTS & 1.8$\pm$0.2 & 0.9$\pm$0.1 & 1.9$\pm$0.4 & 0.8$\pm$0.2\\
Upper Sco TTS and BDs & 0.8$\pm$0.1 & 0.6$\pm$0.1 & ... & ...\\
\enddata
\tablecomments{The slopes ($m$) and intercepts ($b$) reported here correspond to the following linear relationship: log$(M_{dust}/M_{\oplus})$ = $m$ x log$(M_*/M_{\odot})$ + $b$.}
\tablenotetext{a}{These values are the fit to a mixed sample: our modeled objects plus single objects and non-detections not included in our sample, with masses recalculated from their mm fluxes.}
\end{deluxetable}

\subsubsection{TTS Compared to Brown Dwarfs}
\begin{figure*}
    \gridline{\fig{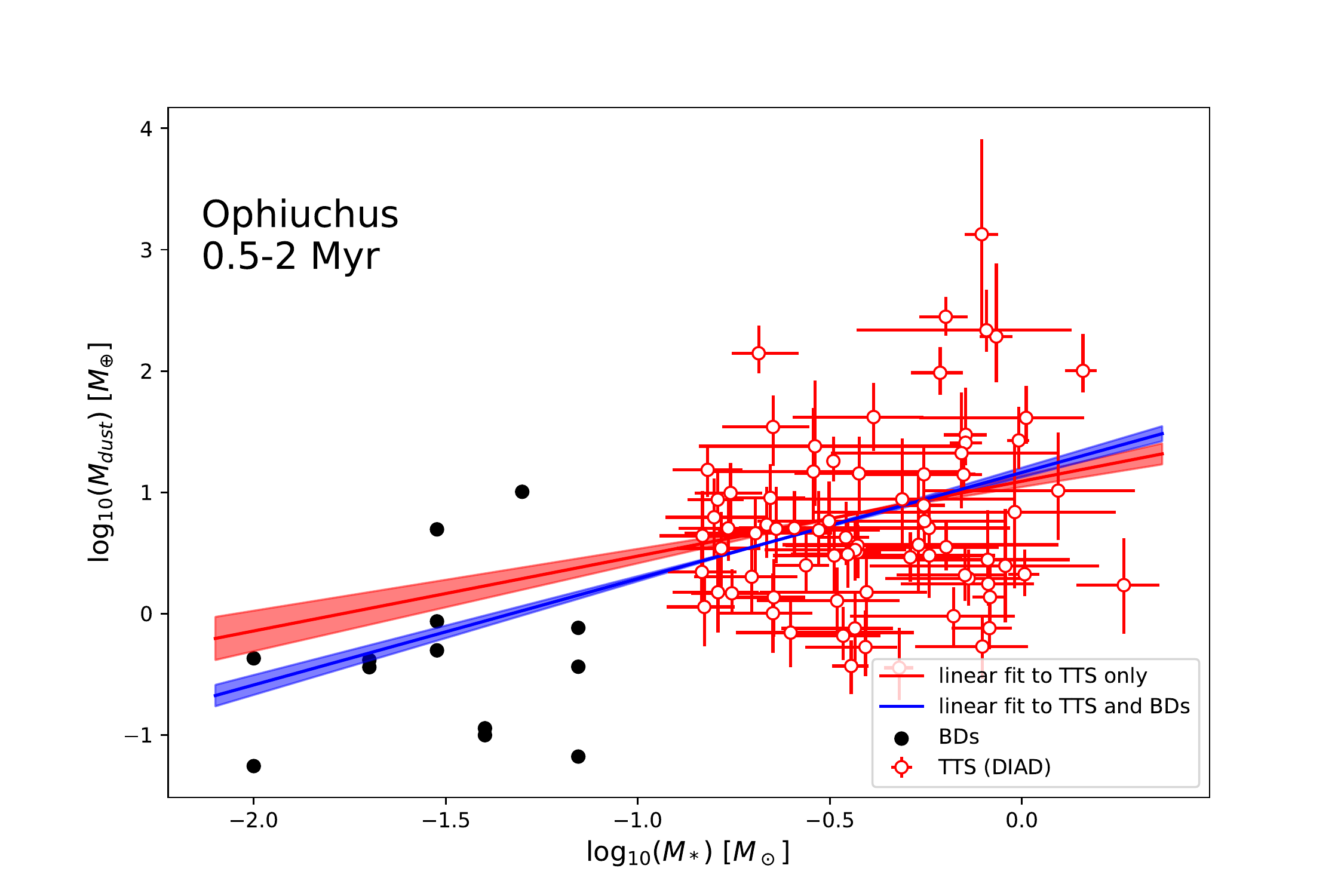}{0.5\textwidth}{(a)}
            \fig{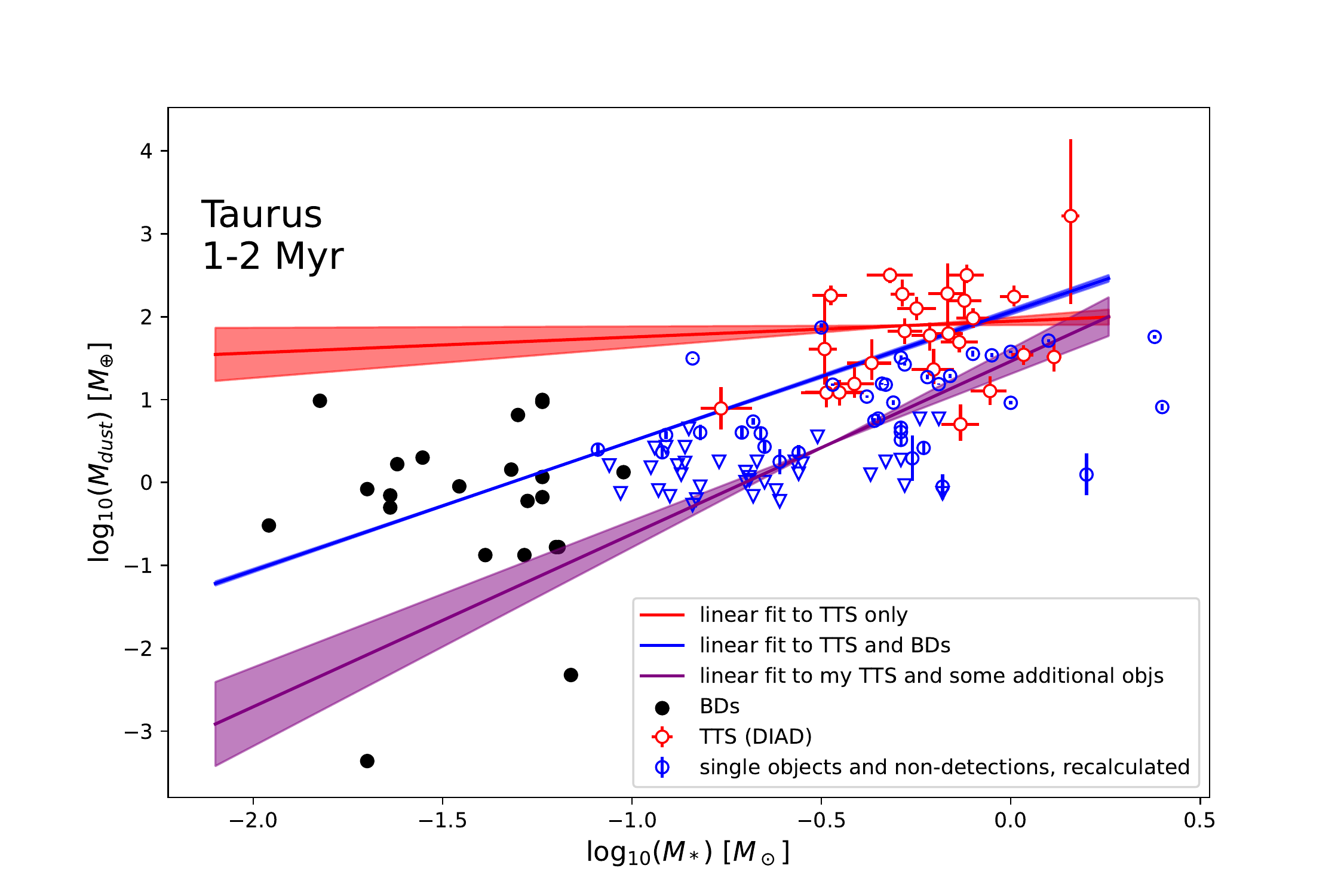}{0.5\textwidth}{(b)}}
    \vspace{-0.5cm}
    \gridline{\fig{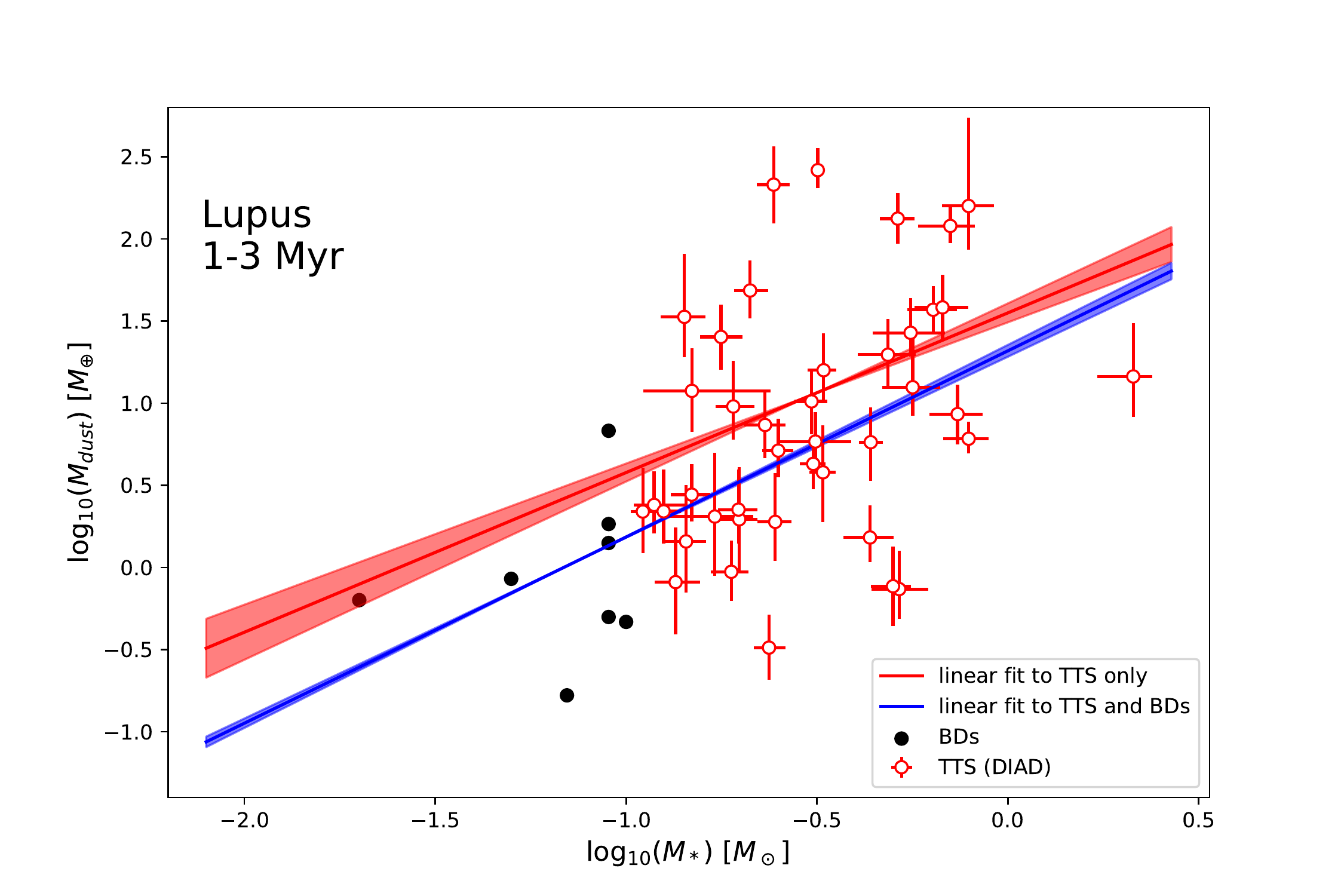}{0.5\textwidth}{(c)}
            \fig{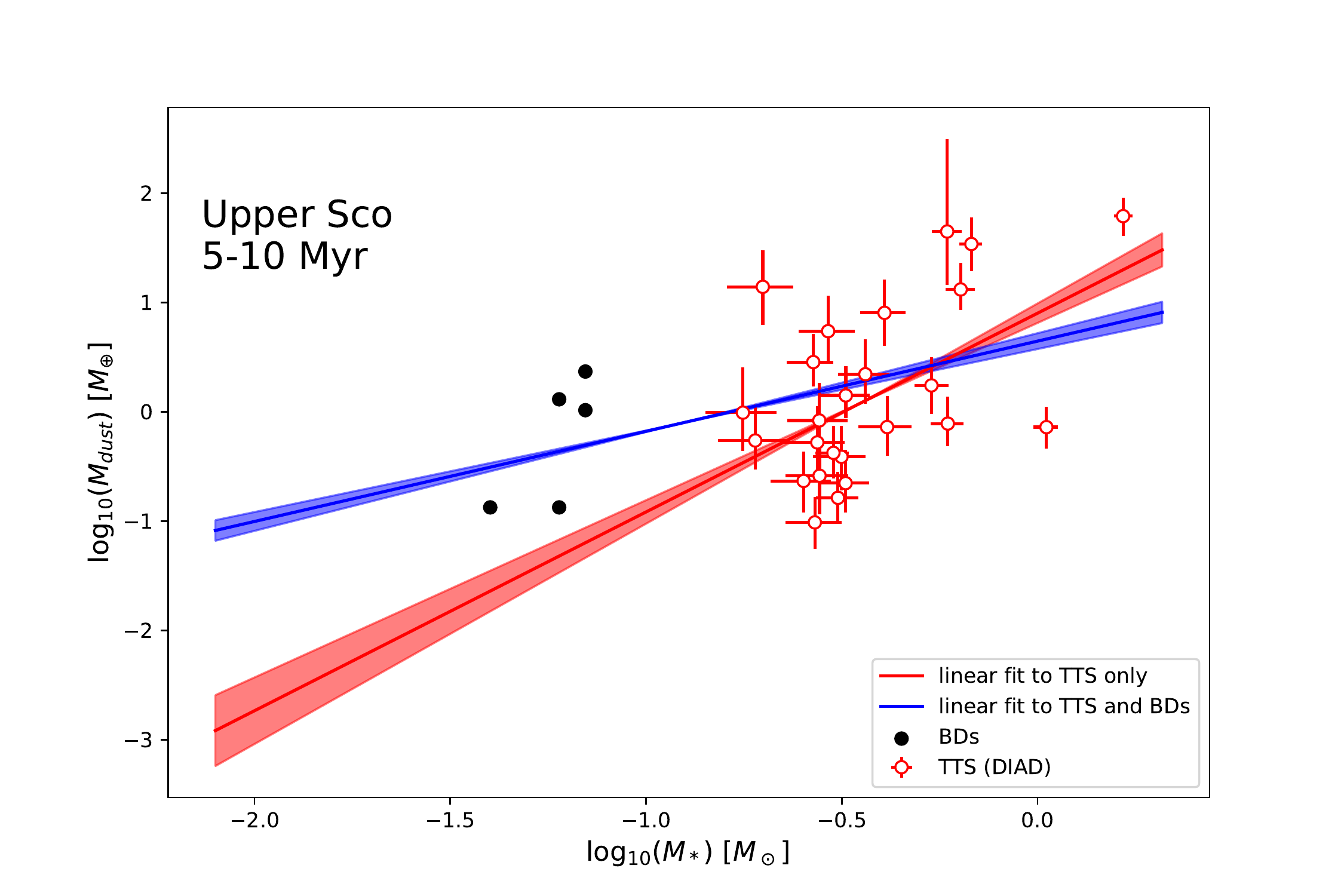}{0.5\textwidth}{(d)}}
    \caption{Disk dust masses plotted versus host masses for the four star-forming regions studied in \citet{rilinger21}. TTS objects are shown as red open circles and BD objects are shown as black filled circles. TTS disk and host masses are from the SED models presented in this work. BD disk masses were derived via SED modeling by \citet{rilinger21} and BD host masses were taken from the literature (see citations in \citet{rilinger21}). Linear fits for just the TTS are represented by red lines; linear fits to the TTS and BDs together are shown in blue. As in Figure \ref{fig:mdiskmstar-pasc}(a), we show linear fits to both our sample and the mixed sample for Taurus.}
    \label{fig:ttsbd}
\end{figure*}

Given the scaling of disk mass with host mass for TTS \citep[e.g.,][and this work]{pascucci16, ansdell16, ansdell17}, we explored whether this trend extends to disks around lower mass brown dwarf (BD, $M_* \lesssim$ 0.08 $M_{\odot}$) hosts. Four of the star-forming regions presented in this work were also studied by \citet{rilinger21}, who used DIAD to obtain SED models for 49 disks around BDs in Ophiuchus, Taurus, Lupus, and Upper Sco. Since the modeling process is consistent between that work and this, we can directly compare the disk masses we infer for TTS to the disk masses \citet{rilinger21} inferred for BDs.

In Figure \ref{fig:ttsbd}, we show how the inclusion of BD disks affects the $M_{dust}$ vs $M_*$ relationships presented above. Using PAIDA, we obtained linear fits for the combined sample of TTS and BDs in each region, which are shown in Figure \ref{fig:ttsbd} by the blue lines. We also reproduce the linear fits to the TTS sample alone from Figure \ref{fig:mdiskmstar-pasc} for comparison. In the three younger regions, the the combined TTS and BD fit is generally in good agreement with the fit to the TTS (see Table \ref{tab:mdiskmstar}). In Upper Sco, however, including the BD disks in the fit results in a significantly shallower slope. In other words, in this region, BD disk dust masses are higher than would be expected from the TTS disk mass -- host mass trend. Instead of decreasing with time (see Section \ref{ages}), BD disk masses appear to remain consistent over a span of $\sim$ 10 Myr.

Studies of protoplanetary disk fractions in various star-forming regions indicate that the fraction of stars that host disks increases for later spectral types \citep{carpenter06, luhmanmamajek12, ribas15, luhman22}. In other words, lower-mass stars appear to retain their disks longer than higher-mass stars. The results presented here add additional evidence to this theory. BD disks are typically less massive than their TTS counterparts in a given star-forming region, as expected based on the established trend of decreasing disk mass with host mass. However, the BD disk dust masses remain constant across regions spanning as much as 10 Myr in age while the corresponding TTS disks exhibit a substantial decrease in disk dust mass (Section \ref{ages}). 

One way in which disks are expected to lose mass is through photoevaporation due to radiation from the central star \citep{ercolano17, weber20}. \citet{picogna21} recently showed that the rate of photoevaporation scales with stellar mass. Given the lower masses of BDs compared to pre-main-sequence stars, photoevaporation rates should be lower in BD disks than TTS disks. That BD disks appear to maintain their disk masses over millions of years, while their TTS counterparts show a decrease in mass, may be the result of slower disk dissipation occurring in these objects. The potential for planet formation in a given disk is thus both enhanced by and inhibited by the mass of its host: a more massive YSO is more likely to host a larger disk early in its lifetime, but it will also likely emit more UV and X-ray radiation, which will dissipate the disk more quickly.

\subsection{Comparison Between Regions of Various Ages}\label{ages}
In this work we studied disks in eleven star-forming regions, which vary in age from $\sim$0.5 to $\sim$10 Myr.  By consistently modeling the SEDs of the disks in each of these regions, we can directly compare their properties and probe how the population of disks changes with time.  In particular, we consider the disk masses and the degree of dust settling; both of these properties have implications for the planet-forming potential of these disks.

\subsubsection{Disk Mass with Age}
\begin{figure*}
    \plotone{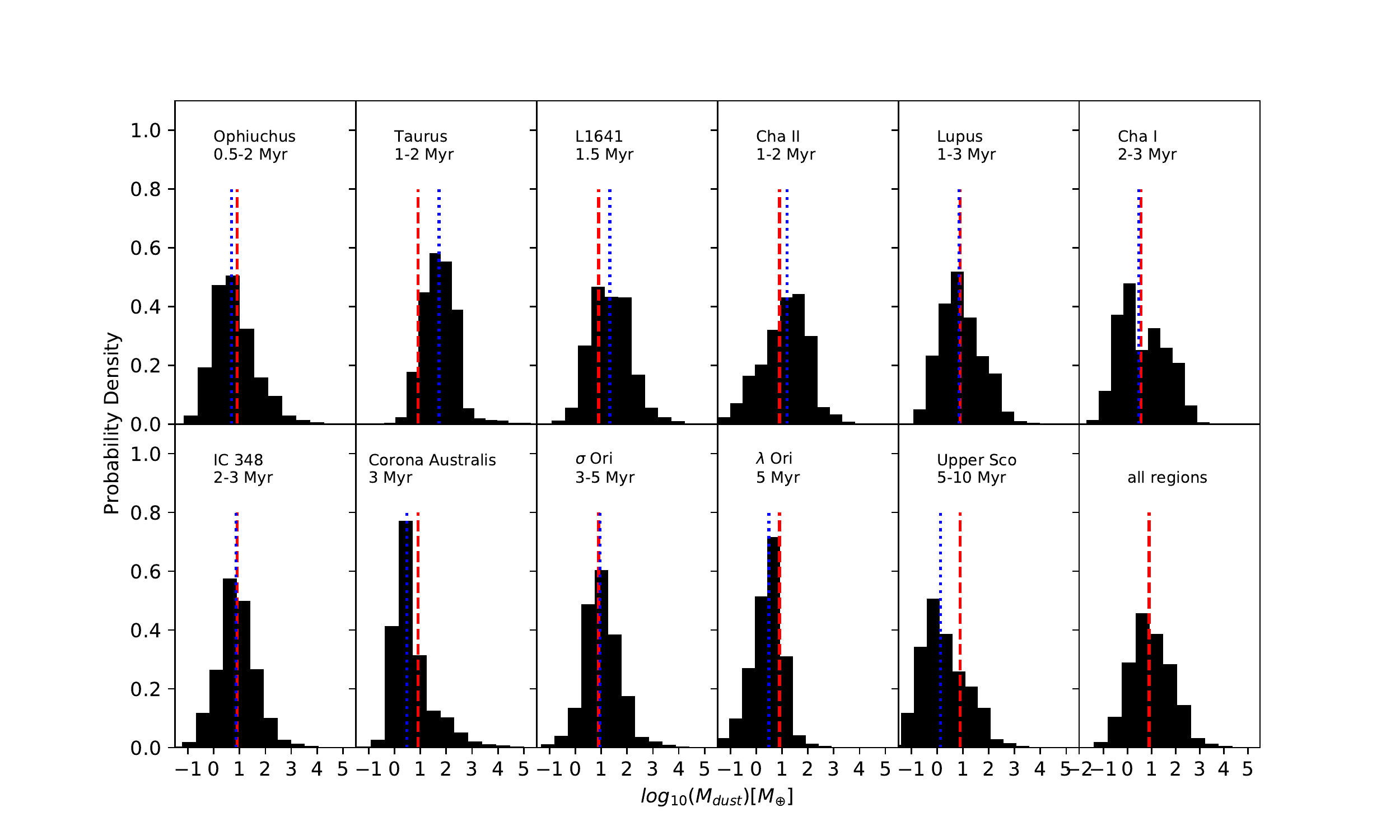}
    \caption{Probability distribution functions of dust mass for each region and for all regions together (lower right panel). The red dashed line in each panel represents the median dust mass for the overall sample ($\sim$ 8 $M_{\oplus}$) and the blue dotted lines represent the median dust mass in each region.}
    \label{fig:masshists}
\end{figure*}

Disk mass distributions for each object are constructed using 10000 random samples from the object's MCMC chain. We use these masses to construct normalized histograms of disk mass for each region. These normalized histograms are plotted in Figure \ref{fig:masshists} for each region, as well as the normalized overall mass distribution including all regions. As shown in Figure \ref{fig:masshists}, the peak of the mass distribution tends to shift to lower masses as the age of the region increases.  Young regions such as Taurus have mass distributions that peak above the median dust mass of the entire sample (represented by the red dashed line in Figure \ref{fig:masshists}), while older regions such as Upper Sco have mass distributions that peak below the median dust mass. The mass distributions of moderately-aged regions such as IC 348 peak near the overall median. 

Two regions appear to deviate from this trend: Ophiuchus and Corona Australis. Ophiuchus has a mass distribution that peaks near the median of the overall sample, despite being the youngest region in our sample. This result is consistent with the findings of \citet{williams19}, who report Ophiuchus disk dust masses to be smaller than the masses of disks in other, slightly older regions. A possible explanation suggested by \citet{williams19} is that the local cloud environment may significantly affect disk masses. \citet{kuffmeier20}, for example, report that star-forming regions with higher ionization rates may contain less massive disks, since ionization can decrease disk size. The mass distribution for disks in Corona Australis also peaks at lower disk masses than may be expected for a moderately young star-forming region. Another potential explanation for the low disk masses in these regions is descibed by \citet{testi22}. Dynamical modeling by \citet{turrini12, turrini19, gerbig19} indicates that formation of planets within a disk can trigger collisional fragmentation of planetesimals, increasing the total mass of grains detectable at millimeter wavelengths. If planets can form within the first 0.5 -- 1 Myr of a disk's lifetime, the formation of this secondary population of dust grains may not have started yet in the disks in Ophiuchus and Corona Australis. Furthermore, recent studies of Corona Australis \citep{cazzoletti19, esplin22} suggest that the region may be comprised of two populations: the young ($<$ 3 Myr) Coronet Cluster, and an older, more extended population. The lower disk masses we find in this region support the existence of this older population. 

Our finding that disk dust masses generally decrease as the age of the region increases is consistent with previous studies \citep[e.g.,][]{ansdell16, barenfeld16, pascucci16, cieza19, vanterwisga19, villenave21, vanterwisga22}, which have found that older regions typically contain less massive disks than their younger counterparts. However, an important caveat is that we have only included objects which have millimeter-wavelength detections.  The different ALMA surveys that studied each region had a range of sensitivities and thus different detection limits; regions observed with lower sensitivities will thus be biased towards higher masses, since the lower mass disks are likely to be less luminous, smaller in radius, and harder to detect. In particular, the survey of $\sigma$ Ori performed by \citet{ansdell17} was only sensitive to dust masses down to $\sim$2 $M_{\oplus}$.  This upper limit is much higher than that achieved by surveys of other regions: \citet{barenfeld16}, \citet{pascucci16}, \citet{ansdell16}, and \citet{ansdell20} all detect dust masses on the order of a few tenths of an Earth mass for Upper Sco, Cha I, Lupus, and $\lambda$ Ori, respectively.  The lower sensitivity of the $\sigma$ Ori survey compared to that of other regions means that this region's mass distribution is biased towards higher dust masses.

\begin{figure*}
    \plotone{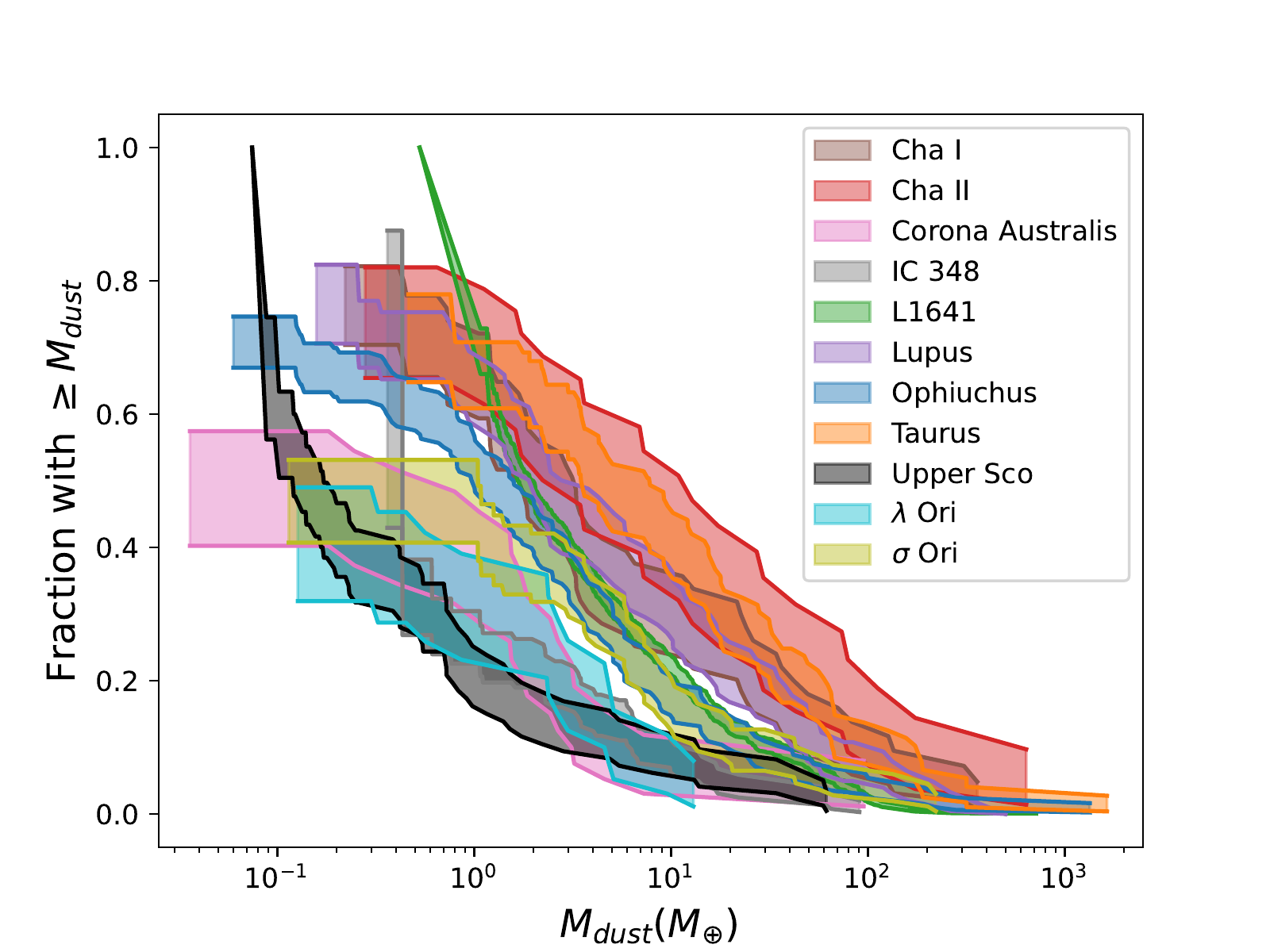}
    \caption{Cumulative distribution functions of dust mass for each region. We include dust masses from our SED modeling for objects in our sample, as well as consistently-recalculated dust masses (see Section \ref{masscomp}) for objects not included in our sample (including non-detections) in order to more fairly compare between regions.}
    \label{fig:lifelines}
\end{figure*}

In order to account for the different sensitivities of the ALMA surveys, we show cumulative distributions of the dust mass in each region in Figure \ref{fig:lifelines}. These distributions were constructed using a Kaplan-Meier esitmator from the \texttt{lifelines} package \citep{davidson-pilon21}\footnote{https://lifelines.readthedocs.io/}. In order to ensure a fair comparison between regions, we include masses for all single objects in each region. We do not include any binary objects or objects in multiple systems, since the presence of a companion can significantly affect the evolution of a disk \citep[e.g.,][]{akeson19, barenfeld19, kounkel19, zurlo20, zurlo21, offner22}. For objects in our sample, we use the masses derived from DIAD. For single objects that were detected in the millimeter but either unable to be fit or not included in our sample, we use the dust mass calculated from the flux measurement (as in Section \ref{masscomp}). For non-detections, we compute the upper limit fluxes as three times the uncertainty plus any positive measured flux density, as in \citet{barenfeld16}; mass upper limits were calculated from these flux upper limits. The L1641 survey targets were selected based on \textit{Herschel} 70 \mum{} detections, which biases the sample towards brighter, more massive disks as noted by \citet{grant21}. To counteract this bias we include objects from the more complete Survey of Orion Disks with ALMA (SODA) project \citep{vanterwisga22}. Even when accounting for the sample biases and non-detections, the results reported above still hold: Figure \ref{fig:lifelines} shows a decrease of disk mass with age, and Ophiuchus and Corona Australis still appear to have lower disk masses than expected from their ages.

\begin{figure}
    \epsscale{1.2}
    \plotone{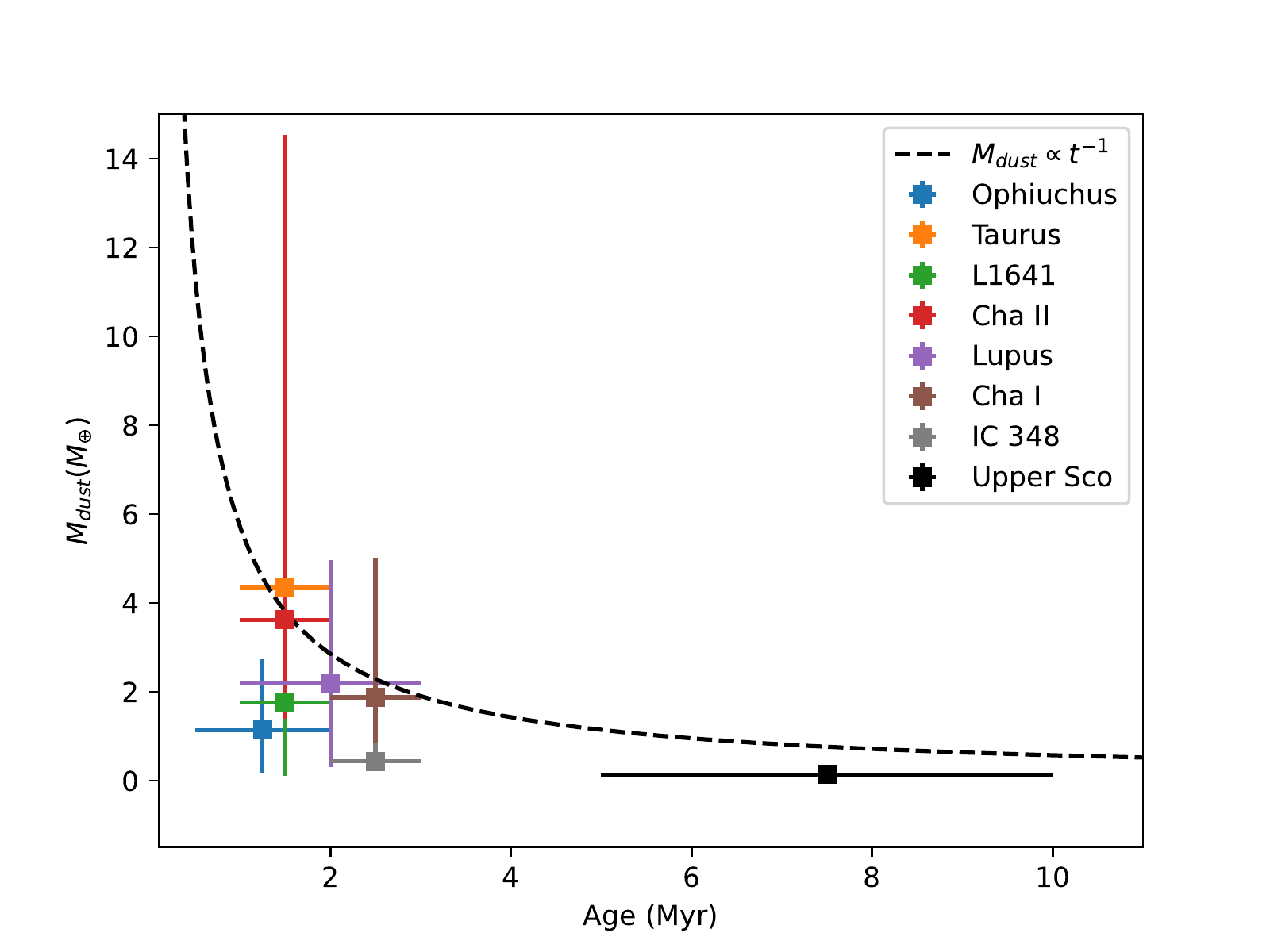}
    \caption{Median disk dust masses for regions with $>$50\% mm-flux detections, plotted versus age of the region. The black dashed line shows the relationship $M_{dust} \propto t^{-1}$, scaled to the weighted mean mass at 1.5 Myr.}
    \label{fig:medsbyage}
\end{figure}

To further explore the decrease in disk mass with time, we plot in Figure \ref{fig:medsbyage} the median disk mass for each region versus age of the region. Median values are determined from the cumulative distribution functions shown in Figure \ref{fig:lifelines} by taking the value of $M_{dust}$ at which 50\% of disks have masses $\geq M_{dust}$. We do not include Corona Australis, $\lambda$ Ori, or $\sigma$ Ori since these regions have $<$ 50\% detections in the millimeter. Following \citet{testi22}, we also show the trend $M_{dust} \propto t^{-1}$. \citet{testi22} normalized this trend to the median dust mass in Ophiuchus. As we have argued above, disks in Ophiuchus tend to have unusually low disk masses for their age; we opt instead to scale the $M_{dust} \propto t^{-1}$ trend to the weighted average median dust mass at 1.5 Myr (i.e., the weighted average of the median values for Taurus, Cha II, and L1641). This trend agrees reasonably well with our median disk masses and indicates that the disk dust mass will be reduced to $\sim$ one-third of its initial amount after 3 Myr, which is consistent with typical disk dissipation timescales \citep{mamajek09, ribas14}.


\subsubsection{Dust Settling with Age}
\begin{figure*}
    \plotone{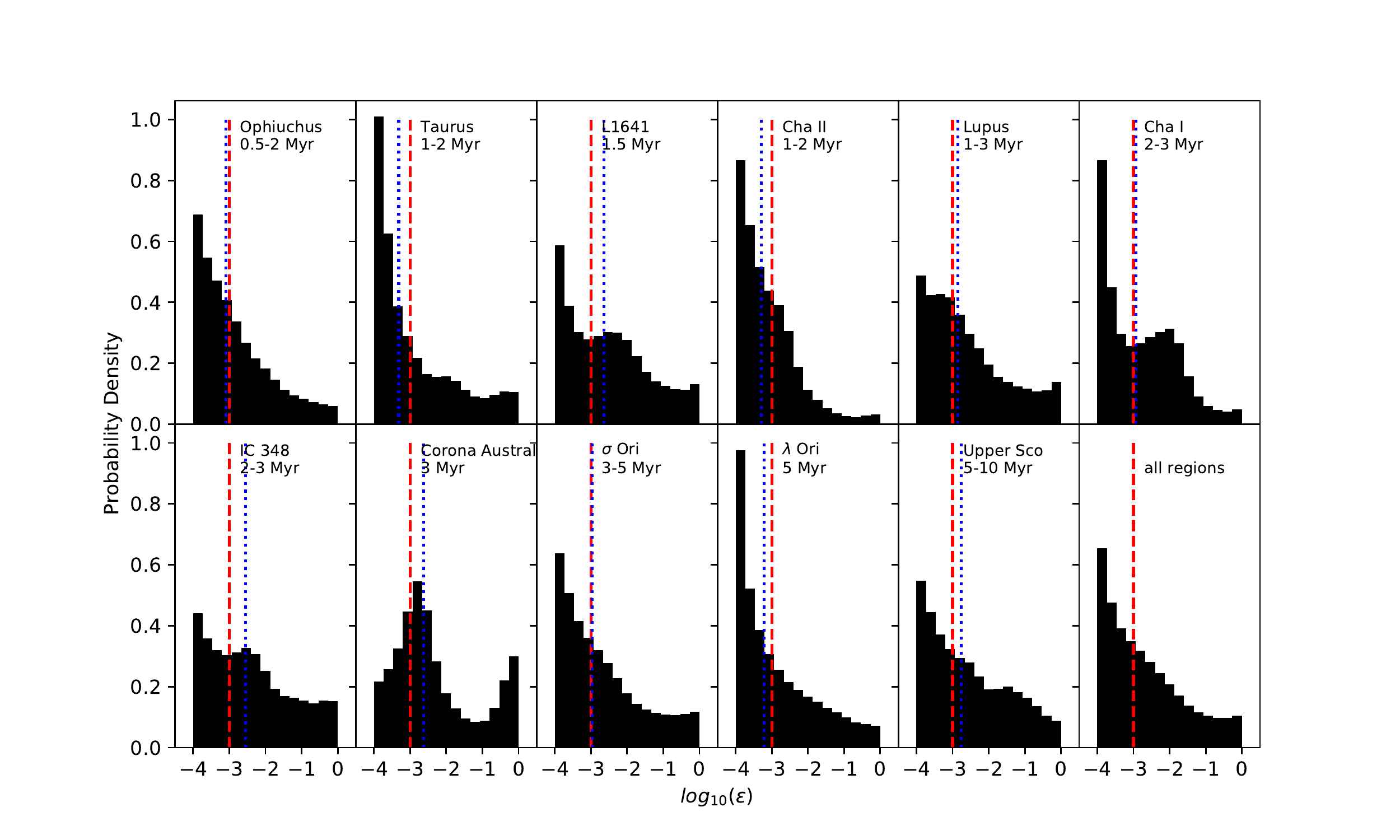}
    \caption{Probability distribution functions of the dust settling parameter $\epsilon$ for each region, and for all regions together (lower right panel). The red dashed line in each panel represents the median $\epsilon$ value for the overall sample ($\sim 10^{-3}$) and the blue dotted lines represent the median $\epsilon$ values for each region.}
    \label{fig:epshists}
\end{figure*}

Figure \ref{fig:epshists} shows the settling parameter $\epsilon$ probability distribution functions for each of the eleven regions as well as the overall distribution for all regions together. Larger values of $\epsilon$ correspond to lower degrees of settling. As for the disk dust mass distributions, these normalized histograms were constructed from 10000 random $\epsilon$ values drawn from the MCMC chain for each object.

The $\epsilon$ distributions for all regions are remarkably similar. Nearly every region peaks at the lowest $\epsilon$ value allowed by our SED modeling, though some regions have a second, smaller peak at higher $\epsilon$ values. We note that $\epsilon$ is defined as the ratio between the dust-to-gas mass ratio in the disk atmosphere and the global dust-to-gas mass ratio of the disk (assumed to be 0.01). Lower values of $\epsilon$ thus correspond to dust-depleted disk atmospheres caused by a high degree of settling. Our observed trend indicates that significant amounts of dust settling has occurred in these disks, regardless of their age. One may expect the degree of settling in a disk to increase with the age of the disk.  However, our result is in agreement with recent studies, which have shown that disks show evidence of dust settling even at young ages \citep[$\sim$ 1Myr;][]{furlan09, lewis16, grant18, rilinger21}.

Large dust grains are expected to settle to the disk midplane faster than small dust grains, since larger grains decouple more easily from the gas in the disk, and feel a drag force as they orbit the central star \citep{dullemond04}.  The high degrees of settling that we find in young disks may therefore indicate that the dust grains in these disks grew significantly in the first $\sim$1 Myr.  Alternatively, high degrees of settling may be the result of low turbulence in these disks.  Turbulence causes the dust and gas in a disk to mix, which counteracts the settling process. If the gas density is sufficiently low \citep{dullemond04}, or the gas velocity in the disk atmosphere is low \citep[e.g.,][]{ciesla07, flaherty15, flaherty17}, turbulence is less efficient and thus dust settling can occur more easily.  The high degrees of settling we find in the disks studied here may therefore be the result of rapid dust grain growth and/or lack of turbulence in the upper layers of the disks.

\section{Summary and Conclusion}\label{summary}
Using the physically motivated DIAD models in conjunction with an ANN, we have obtained SED fits for 338 disks around T Tauri stars. These disks are located in the following eleven star-forming regions, ranging in age from $\sim$ 0.5 -- 10 Myr old: Ophiuchus, Taurus, L1641, Cha II, Lupus, Cha I, IC 348, Corona Australis, $\sigma$ Ori, $\lambda$ Ori, and Upper Scorpius. Our main results are as follows:
\begin{enumerate}
\item We confirm that the previously-reported mm-flux-based masses are a factor of 1.5-5 lower than the masses derived from physical models, consistent with results in \citet{ballering19, ribas20}. Masses derived from millimeter fluxes depend on the assumption that the disk is optically thin at that wavelength; if the disk is optically thick, this assumption results in an underestimate of the disk mass. Our SED models do not rely on this assumption, and thus yield more accurate masses. The discrepancy between the two methods is higher for the more massive disks when the disk temperature is scaled by the stellar luminosity. We present two equations (Equations \ref{mcorr1} and \ref{mcorr2}) to be used for correcting millimeter flux-derived disk masses to account for this effect.
\item We find $M_{dust}$ -- $M_*$ relationships that are generally consistent with those reported in \citet{pascucci16}. Five of the star-forming regions studied here (Taurus, Lupus, Cha I, $\sigma$ Ori, and Upper Sco) show statistically significant (p $<$ 0.05), moderate correlations. Ophiuchus, L1641, and IC348 do not have a statistically significant trend; Cha II, Corona Australis, and $\lambda$ Ori have too few points to accurately assess their $M_{dust}$ -- $M_*$ relationships. We do not find any trend of this relationship with age.
\item Brown dwarfs are generally consistent with the $M_{dust}$ -- $M_*$ relationships observed for TTS in younger regions. In the 5-10 Myr-old Upper Sco region, however, brown dwarf disks are more massive than predicted from their host masses. This result may indicate that brown dwarf disks dissipate more slowly than their higher-mass companions; photoevaporation is more efficient in higher-mass stars, which may explain this trend.
\item We find clear evolution in the disk masses with time, with older regions having lower disk masses than younger regions, in agreement with previous studies \citep[e.g.,][]{barenfeld16, vanderplas16, vanterwisga22}. As an important caveat, different sensitivities of ALMA surveys of various star-forming regions may have biased portions of our sample towards higher masses.  Future, more sensitive surveys may reveal a larger population of low-mass disks, especially in older regions.
\item The degree of dust settling appears to be consistent across age, with even the youngest regions showing appreciable settling. This result agrees with \citet{ribas17} and \citet{grant18}, who also reported appreciable levels of dust settling in regions as young as 1 Myr.
\end{enumerate}

In conclusion, these results may help to ease the reported ``missing'' mass problem.  By assuming the disks to be optically thin, previous studies may have underestimated the masses of disks around T Tauri stars; our physical models are not limited by this assumption and yield masses that are more consistent with observed planetary systems.  Furthermore, the high degrees of dust settling we find in disks of all ages may indicate that dust processing and evolution happens quickly; thus planetary systems may also form earlier than previously theorized.

\vspace{\baselineskip}
We thank the referee for the careful reading of the manuscript and helpful feedback. AMR thanks A. Meredith Hughes and Philip Muirhead for their insightful comments and discussion. This work was funded by NASA ADAP 80NSSC20K0451. We acknowledge support from the NRAO Student Observing Support program through award SOSPA8-007. Á.R. has been supported by the UK Science and Technology research Council (STFC) via the consolidated grant ST/S000623/1 and by the European Union’s Horizon 2020 research and innovation programme under the Marie Sklodowska-Curie grant agreement No. 823823 (RISE DUSTBUSTERS project).This work has made use of data from the European Space Agency (ESA) mission {\it Gaia} (\url{https://www.cosmos.esa.int/gaia}), processed by the {\it Gaia} Data Processing and Analysis Consortium (DPAC, \url{https://www.cosmos.esa.int/web/gaia/dpac/consortium}). Funding for the DPAC has been provided by national institutions, in particular the institutions participating in the {\it Gaia} Multilateral Agreement. This paper utilizes the D’Alessio Irradiated Accretion Disk (DIAD) code. We wish to recognize the work of Paola D’Alessio, who passed away in 2013. Her legacy and pioneering work live on through her substantial contributions to the field.

\appendix
\section{Agreement with Reported  Literature Values}\label{app:tstarrstar}
\renewcommand{\thefigure}{A.\arabic{figure}}
\setcounter{figure}{0}

Here we present a comparison between the stellar values obtained from our SED modeling and values reported in the literature. As described in Section \ref{ann}, we incorporate literature values as priors for stellar parameters when literature values exist, and we ensure our $T_*$ and $R_*$ values are consistent with stellar evolution models using MIST \citep{paxton11, paxton13, paxton15, dotter16, choi16}. For a given $M_*$ and $Age_*$, $T_*$ and $R_*$ are calculated with MIST; these four values are then used in our MCMC fitting process.

Literature values for $M_*$, $T_*$ and $R_*$ are taken from: \citet{walter97, luhman99, lopez-marti05, wilking05, lada06, cieza07, forbrich07, luhman07, muench07, luhman08, sicilia-aguilar08, meyer09, cieza10, hernandez10, mcclure10, bayo11, erickson11, rigliaco12, andrews13, spezzi13, alcala14, herczeg14, hernandez14, manara15, barenfeld16, manara16a, manara16, pascucci16, alcala17, ansdell17, grant18, ruiz-rodriguez18, cazzoletti19, ansdell20}. Stellar temperatures are determined from spectral types; for objects with reported spectral types but no reported temperatures, we used the \citet{pecaut13} scaling to convert the reported spectral types to $T_*$ values. Stellar radii were either obtained via comparison to a reference SED \citep{spezzi08, spezzi13}, or calculated from the stellar luminosity and $T_*$. Luminosities were obtained by applying a bolometric correction to the observed $J$ band magnitude.

\begin{figure*}
    \gridline{\fig{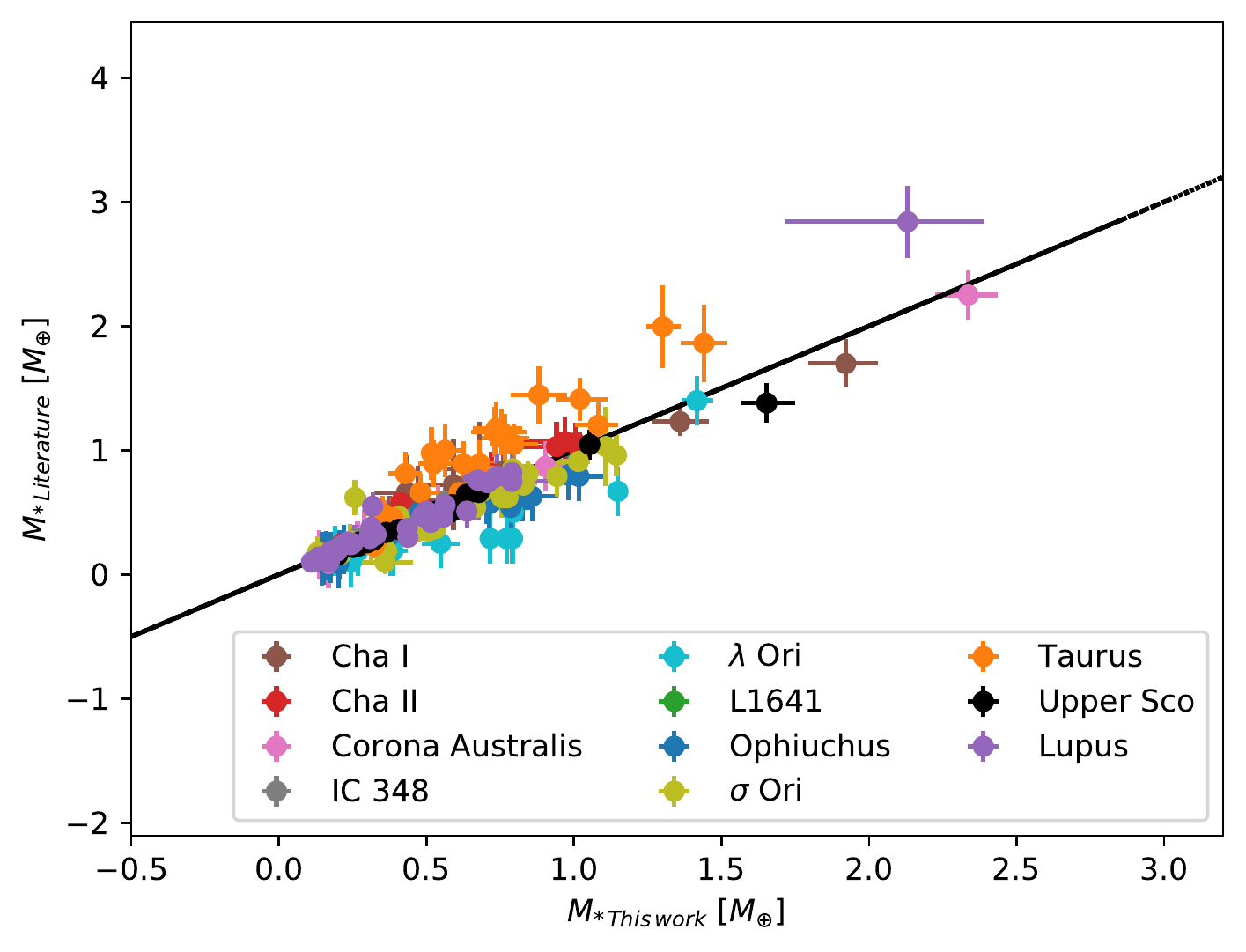}{0.33\textwidth}{(a)}
            \fig{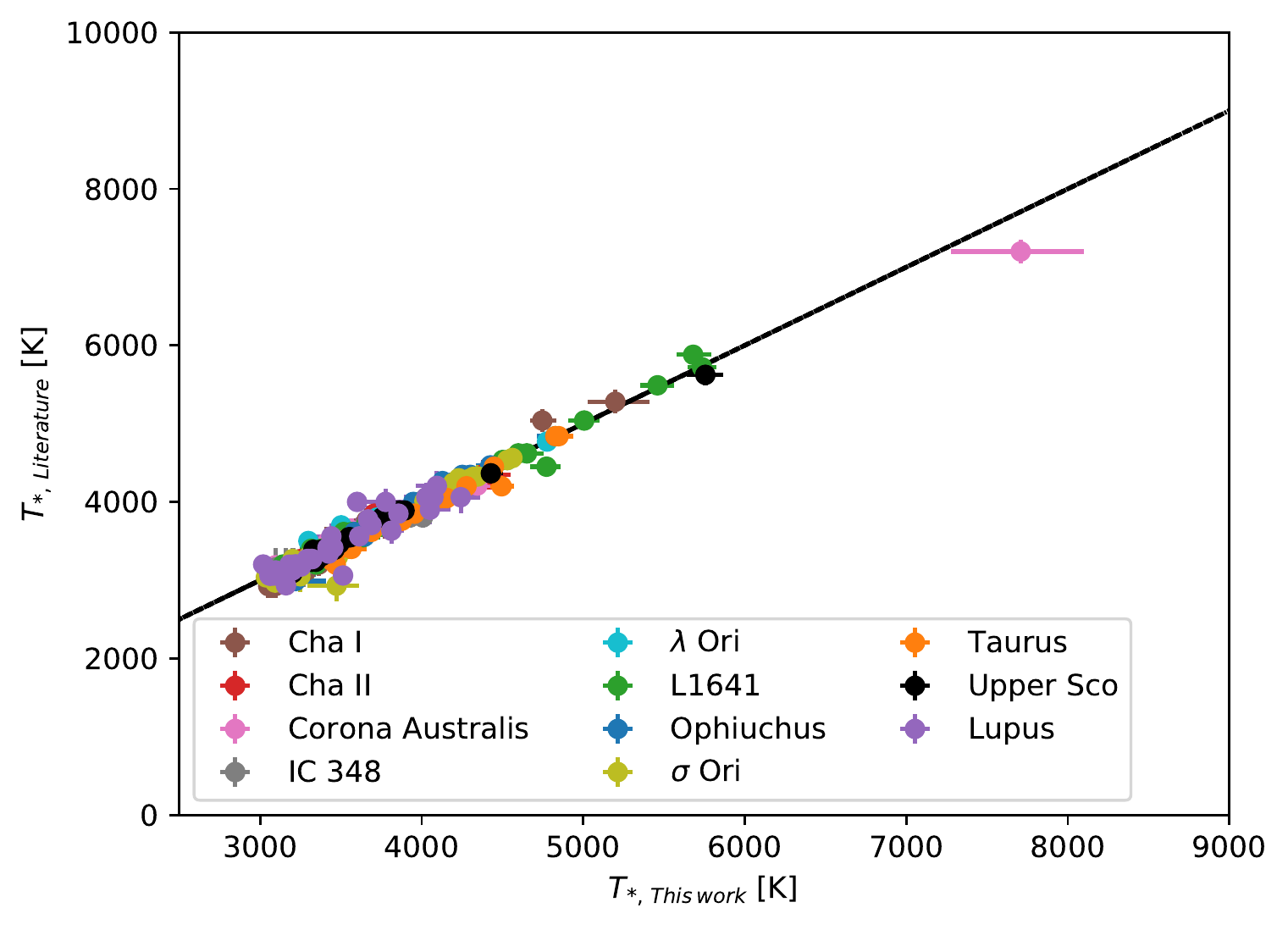}{0.33\textwidth}{(b)}
            \fig{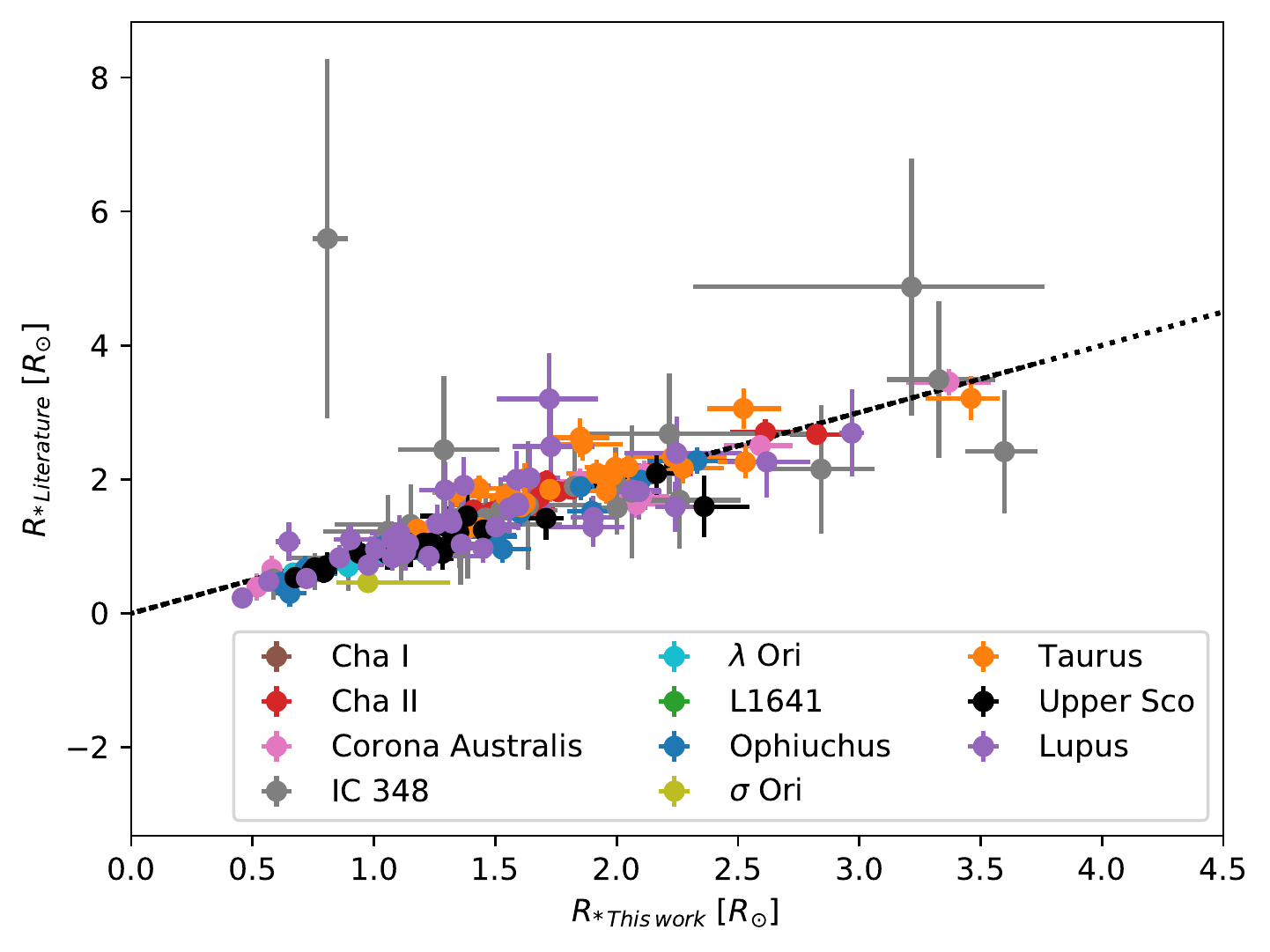}{0.33\textwidth}{c}}
    \caption{Comparison between reported values and values obtained via our SED modeling for $M_*$ (panel a), $T_*$ (panel b) and $R_*$ (panel c). The black dotted line in each panel represents a one-to-one correlation. Our modeled values for each of these parameters agree very well with previously reported values.}
    \label{fig:tstarrstar-lit}
\end{figure*}

\begin{figure}
    \plotone{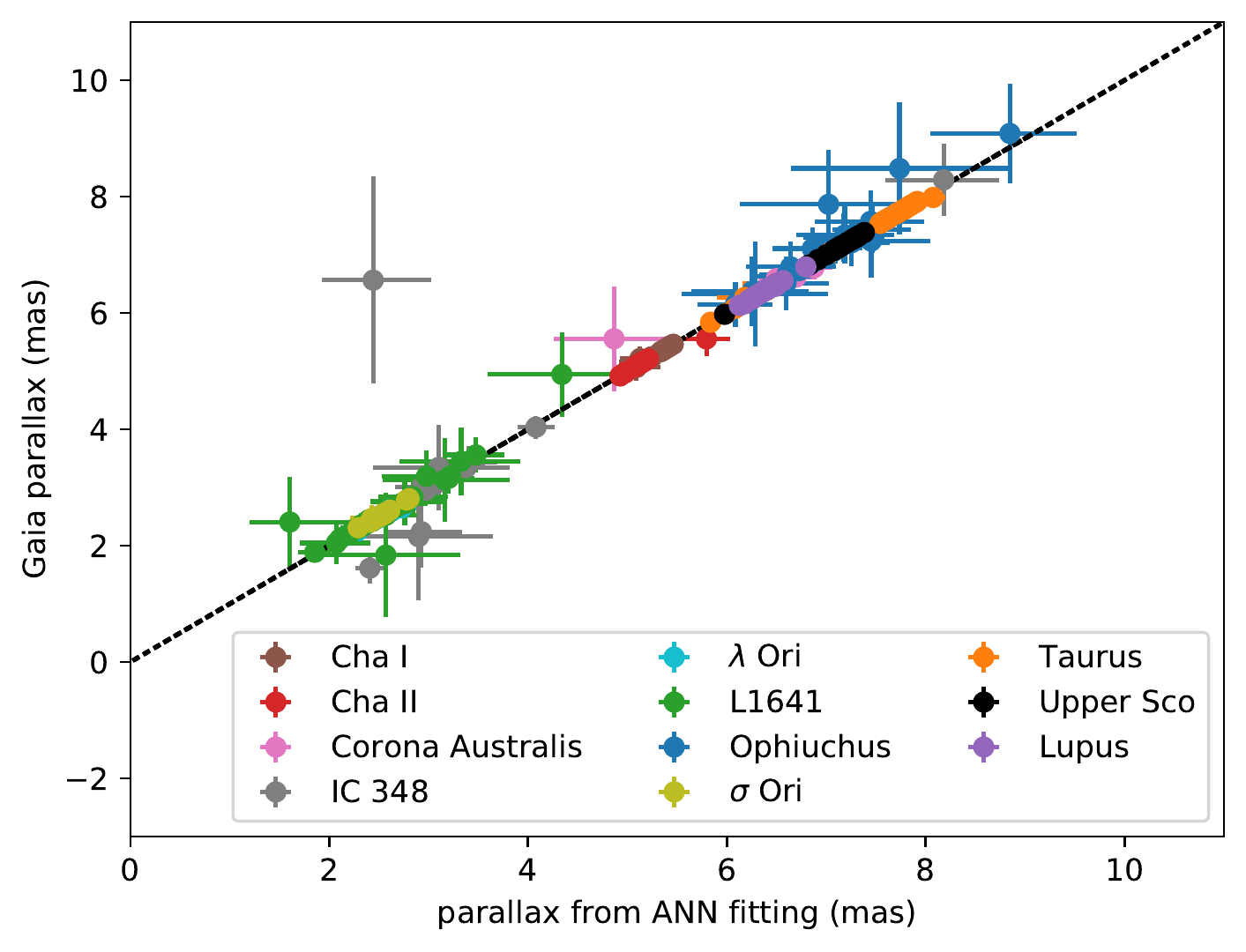}
    \caption{Comparison between parallax values obtained via our SED modeling and those measured by Gaia. The black dotted line represents a one-to-one correlation. Our parallax values agree very well with previously reported values.}
    \label{fig:gaiacomp}
\end{figure}

In Figure \ref{fig:tstarrstar-lit}, we compare the previously reported $M_*$, $T_*$ and $R_*$ values to those obtained via our SED modeling. For all three parameters, we find very strong agreement. We also present a comparison of the best-fit parallaxes we obtain for each object versus the parallaxes reported by Gaia in Figure \ref{fig:gaiacomp}, which also shows a strong agreement. As mentioned in Section \ref{ann}, Gaia parallaxes are included as a prior in our model, so a strong correlation is expected.

\begin{figure}
    \plotone{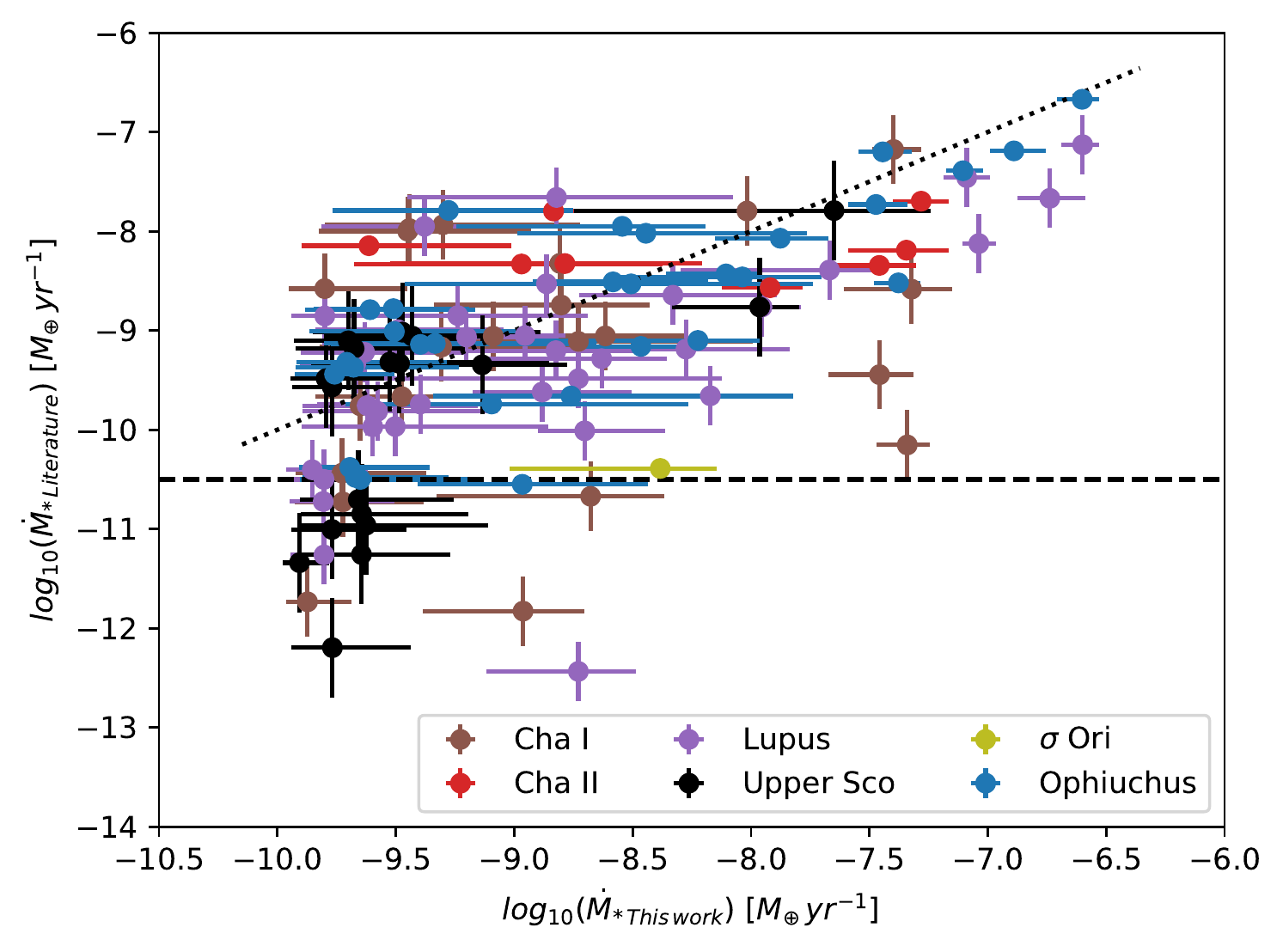}
    \caption{Comparison between mass accretion rate values obtained via our SED modeling and those reported in the literature. The black dotted line represents a one-to-one correlation. The black dashed line denotes the approximate threshold below which chromospheric activity interferes with derivations of \dotM{} in the literature \citep{manara13}.}
    \label{fig:mdotcomp}
\end{figure}

Figure \ref{fig:mdotcomp} shows a comparison between our modeled mass accretion rates (\dotM{}) and \dotM{} values reported in the literature (scaled to Gaia distances). Literature \dotM{} values are taken from \citet{natta06} for Ophiuchus, \citet{antoniucci11} for Cha II, \citet{alcala17} for Lupus, \citet{manara17} for Cha I, \citet{natta14} for Sigma Ori, and \citet{manara20} for Upper Sco. These works calculate the accretion luminosity, and hence \dotM{}, from Br$\gamma$, Pa$\beta$ and other emission line equivalent widths \citep{natta06, antoniucci11, natta14} or UV excess \citep{alcala17, manara17, manara20}. We note that the ANN was trained on $log_{10}$(\dotM{}) values between -10 and -6.5. Some objects have reported \dotM{} values outside this range, though these objects have $\alpha$ values that are well-constrained, so this does not introduce a bias in our disk masses.

The \dotM{} error bars from DIAD are generally much larger than those obtained by measuring \dotM{} with emission lines or UV excess. In principle these two accretion rates do not necessarily need to match, since the mass accretion rate onto the star could be different than in the disk. The mass accretion rate onto the star could also be variable \citep[e.g.,][]{robinson19}, by as much as two orders of magnitude \citep{claes22}, while the mass accretion rate probed by DIAD would be an average that sets the surface density profile in an alpha-disk. Finally, we note that chromospheric emission can create noise in UV observations, making it difficult to derive accurate \dotM{} values below the noise threshold \citep{ingleby11}. \citet{manara13} report a noise threshold of $\sim 10^{-10.5}\ \msunyr{}$ for objects with similar $M_*$ and ages as the objects presented here. Despite these challenges, we find a general agreement between \dotM{} values from our modeling and \dotM{} values from the literature above the noise threshold.

\bibliographystyle{apj}
\bibliography{refs}

\end{document}